\documentclass[fleqn,usenatbib]{mnras}
\usepackage[T1]{fontenc}
\usepackage{graphicx} 
\usepackage{amsmath}
\usepackage{amssymb}
\usepackage{color}
\usepackage{multicol}
\usepackage{threeparttable}
\usepackage{newtxtext,newtxmath}

\DeclareRobustCommand{\VAN}[3]{#2}
\let\VANthebibliography\thebibliography
\def\thebibliography{\DeclareRobustCommand{\VAN}[3]{##3}\VANthebibliography}

\def\refjnl#1{{\rm#1}}
\def\apjl{\refjnl{ApJ}}                
\def\apjs{\refjnl{ApJS}}               
\def\ao{\refjnl{Appl.~Opt.}}           
\def\aap{\refjnl{A\&A}}                

\def\be{\begin{equation}} 
\def\ee{\end{equation}}

\def\msun{{\Msun}}

\def\HI{\hbox{H~$\scriptstyle\rm I\ $}} 
\def\HII{\hbox{H~$\scriptstyle\rm II\ $}}

\def\gsim{\lower.5ex\hbox{\gtsima}} 
\def\lsim{\lower.5ex\hbox{\ltsima}} \def\gtsima{$\; \buildrel > \over 
\sim \;$} \def\ltsima{$\; \buildrel < \over \sim \;$} \def\prosima{$\; 
\buildrel \propto \over \sim \;$} \def\gsim{\lower.5ex\hbox{\gtsima}} 
\def\lsim{\lower.5ex\hbox{\ltsima}} 
\def\simgt{\lower.5ex\hbox{\gtsima}} 
\def\simlt{\lower.5ex\hbox{\ltsima}} 
\def\simpr{\lower.5ex\hbox{\prosima}} \def\la{\lsim} \def\ga{\gsim} 
  
 \def\gtsima{$\; \buildrel > \over \sim \;$} 
\def\ltsima{$\; \buildrel < \over \sim \;$} 
\def\gsim{\lower.5ex\hbox{\gtsima}} 
\def\lsim{\lower.5ex\hbox{\ltsima}} 
\def\simgt{\lower.5ex\hbox{\gtsima}} 
\def\simlt{\lower.5ex\hbox{\ltsima}} 
\def\simpr{\lower.5ex\hbox{\prosima}} 
\def\la{\lsim} 
\def\ga{\gsim}

\def\msun{\,{\rm \Msun}}

\def\E3{{\cal E}_{\rm g}^{III}}

\def\Msun{\rm M_\odot}

\def\Msun{\rm M_\odot}

\def\M*{M_*}
\def\Z*{Z_*}
\def\L*{L_*}

\title[Sources of reionization]{Astraeus III: The environment and physical properties of reionization sources} 
\author[Hutter et al.]{Anne Hutter$^{1}$\thanks{a.k.hutter@rug.nl}, Pratika Dayal$^1$, Laurent Legrand$^1$, Stefan Gottl\"ober$^2$, Gustavo Yepes$^{3,4}$\\
$^{{1}}$ Kapteyn Astronomical Institute, University of Groningen, P.O. Box 800, 9700 AV Groningen, The Netherlands \\
$^{2}$ Leibniz-Institut f\"ur Astrophysik, An der Sternwarte 16, 14482 Potsdam, Germany\\
$^{3}$ Departamento de Fısica Teorica, Modulo 8, Facultad de Ciencias, Universidad Autonoma de Madrid, 28049 Madrid, Spain\\
$^{4}$ CIAFF, Facultad de Ciencias, Universidad Autonoma de Madrid, 28049 Madrid, Spain\\
}

\date{Accepted 2021 March 23. Received 2021 March 02; in original form 2020 September 02}

\pubyear{2021}

\begin{document}
\label{firstpage}
\pagerange{\pageref{firstpage}--\pageref{lastpage}}
\maketitle

\begin{abstract}
In this work, we use the {\sc astraeus} (seminumerical rAdiative tranSfer coupling of galaxy formaTion and Reionization in N-body dArk mattEr simUlationS) framework which couples galaxy formation and reionization in the first billion years. Exploring a number of models for reionization feedback and the escape fraction of ionizing radiation from the galactic environment ($f_\mathrm{esc}$), we quantify how the contribution of star-forming galaxies {(with halo masses $M_h>10^{8.2}\msun$)} to reionization depends on the radiative feedback model, $f_\mathrm{esc}$, and the environmental over-density. Our key findings are: (i) for constant $f_\mathrm{esc}$ models, intermediate-mass galaxies (with halo masses of $M_h\simeq10^{9-11}\msun$ and absolute UV magnitudes of $M_{UV} \sim -15$ to $-20$) in intermediate-density regions (with over-density $\log_{10}(1+\delta) \sim 0-0.8$ on a $2$~comoving Mpc spatial scale) drive reionization; (ii) scenarios where $f_\mathrm{esc}$ increases with decreasing halo mass shift the galaxy population driving reionization to lower-mass galaxies ($M_h\lesssim10^{9.5}\msun$) with lower luminosities ($M_{UV} \gtrsim-16$) and over-densities ($\log_{10}(1+\delta) \sim 0-0.5$ on a $2$~comoving Mpc spatial scale); (iii) reionization imprints its topology on the ionizing emissivity of low-mass galaxies ($M_h\lesssim10^{9}\msun$) through radiative feedback. Low-mass galaxies experience a stronger suppression of star formation by radiative feedback and show lower ionizing emissivities in over-dense regions; (iv) a change in $f_\mathrm{esc}$ with galaxy properties has the largest impact on the sources of reionization and their detectability, with the radiative feedback strength and environmental over-density playing a sub-dominant role; (v) JWST-surveys (with a limiting magnitude of $M_{UV} = -16$) will be able to detect the galaxies providing $\sim 60-70\%$ ($\sim 10\%$) of reionization photons at $z=7$ for constant $f_\mathrm{esc}$ models (scenarios where $f_\mathrm{esc}$ increases with decreasing halo mass).

\end{abstract}

\section{Introduction}
\label{sec_introduction}
The Epoch of Reionization marks a pivotal time in the history of our Universe when the photons emitted by the first stars and galaxies permeate and gradually ionize the intergalactic medium (IGM). Increasingly robust constraints on reionization, from quasar absorption spectra and {\it Planck} measurements of the cosmic microwave background (CMB), hint at a scenario where reionization reaches its midpoint around $z\simeq7.5$ and ends at $z\simeq6$ \citep{davies2018, fan2006, planck2018}. Such a scenario is in agreement with star-forming galaxies being the main drivers of reionization with a small contribution from active galactic nuclei \citep[AGN;][]{parsa2018, hassan2018, dayal2020}. Indeed, the observed ultra-violet (UV) luminosity function of star-forming galaxies at $z=6-10$ \citep{bhatawdekar2019, bouwens2019} has been used to infer that those galaxies, extending down to absolute UV magnitudes of $M_{UV} \simeq -13$ \citep{bouwens2015b}, could provide the necessary number of ionizing photons \citep{ishigaki2018}. However, given the complexity of understanding crucial ingredients, such as the minimum halos mass of star forming/black-hole hosting galaxies, the efficiency of star formation/black-hole accretion and the corresponding ionizing photon emissivities, the escape fraction of ionizing photons from early galaxies ($f_\mathrm{esc}$) and the impact of reionization feedback, it remains unclear which (and if only) star-forming galaxies are the key drivers of reionization \citep[see e.g.][]{dayal2018}. In this work, we focus on studying the star-forming galaxies that could be the key drivers of reionization.

The intrinsic ionizing emissivity from galaxies is determined by the star formation efficiency of low-mass halos and the minimum halo mass that can support star formation; observationally, these can be translated into the the faint-end slope and cut-off of the UV luminosity function. These are shaped by two forms of feedback: that from supernovae type II (SNII) explosions and photoheating by reionization. While supernovae (SN) explosions reduce the amount of gas for star formation on timescales of less than $30$~Myr, the photo-heating of gas surrounding a galaxy in an ionized regions leads to a reduction of the gas mass (delayed by a dynamical timescale) and associated star formation in low-mass galaxies. As the pressure of the heated gas rises, an increasing fraction of gas photo-evaporates into the IGM \citep{barkana-loeb1999, shapiro2004} and increases the Jeans mass for galaxy formation gradually \citep{couchman-rees1986, hoeft2006}. Both these feedback mechanisms suppress star formation in low-mass halos, which could limit their reionization contribution despite their high number-densities. 

Further, the fraction of ionizing photons produced within the galaxy that can escape into the IGM, $f_\mathrm{esc}$, can alter the {\it escaping} ionizing emissivity from a galaxy significantly. While direct observations of $f_\mathrm{esc}$ during reionization are impossible due to the absorption of the escaping ionizing radiation by neutral hydrogen (\HI) in the IGM, simulations have shown that $f_\mathrm{esc}$ depends on the physical processes and gas distribution within and around the galaxy. Gas outflows due to radiative feedback \citep{kitayama2004, whalen2004, abel2007}, SN explosions \citep{Kimm2014, Kimm2017, Kimm2019} and mass accretion onto black holes \citep{Trebitsch2018, Seiler2018} carve out low-density tunnels, through which ionizing photons can escape into the IGM. At the same time, as a galaxy's gravitational potential deepens by accreting more mass, it becomes less susceptible to its gas being ejected by these feedback mechanisms. 
Combining these two processes suggests an escape fraction that decreases with increasing halo mass, as has also been shown by a number of simulations \citep{paardekooper2015, Kimm2017, Kimm2019, wise2014}. However, if an increasing star formation rate goes along with a rising porosity of the interstellar medium (ISM) and thus more escape paths for the ionizing photons, the escape fraction could increase with halo mass \citep{wise2009}. This scenario seems to be in line with the observed higher star formation rate densities of Lyman-Continuum (LyC) leaking galaxies \citep{naidu2020}; although it must be noted that some of these galaxies could have non-negligible AGN activity.
This discrepancy between the latest simulation results and observational implications shows that the dependency of the ionizing escape fraction remains still highly uncertain, and with it the galaxy population that drives reionization \citep[see Sec. 7.1][]{dayal2018, Seiler2019}.

Additionally, all the physical processes described above also depend on the location of a galaxy in the large-scale cosmic web structure. The gas accretion and reionization history of a galaxy are crucially shaped by the cosmic gas available around the galaxy and the abundance and masses of its neighbouring galaxies. For example, in case of an inside-out reionization topology, where over-dense regions are ionized before under-dense regions, the radiative feedback from reionization would result in a stronger suppression of star formation in over-dense than in under-dense regions at a given redshift; this would imply lower ionizing emissivities for low-mass galaxies in over-dense regions.

A number of numerical simulations and theoretical models have addressed the question of which galaxies have been the main drivers of reionization. While self-consistent radiation hydrodynamical simulations find low- and intermediate-mass galaxies ($M_h\lesssim10^{10}\msun$) to be the main drivers \citep{lewis2020, katz2018, Kimm2014}, observation-driven, phenomenological models where $f_\mathrm{esc}$ scales with the star formation surface density yield scenarios wherein bright ($M_\mathrm{UV}<-16$) and intermediate to massive ($M_h\gtrsim10^{9.5}\msun$) galaxies are the main contributors to reionization \citep{sharma2016, naidu2020}.

In this paper, we aim at quantifying which physical processes play the dominant role in determining the galaxies that contribute most of the ionizing photons and hence drive reionization. We investigate how SN and radiative feedback or the escape fraction of ionizing photons can redistribute the ionizing photon contribution within the galaxy population, and whether these processes affect the ionizing emissivities of galaxies in different parts of the cosmic web differently, i.e. do galaxies contribute different amounts of ionizing photons depending on whether they lie in over- or under-dense regions.
To answer these questions, we analyse a set of self-consistent, seminumerical simulations of galaxy evolution and reionization that were run within the {\sc astraeus} framework \citep{hutter2021}. This framework explores the entire plausible range of radiative feedback models (ranging from a weak, time-delayed to a strong instantaneous reduction of gas in the galaxy) and the ionizing escape fraction (ranging from being constant for all galaxies at all redshifts to being proportional to the fraction of gas ejected from the galaxy by SN explosions).

This paper is organised as follows. In Section \ref{sec_sims} we briefly describe the physical processes included in the simulations and the different runs. In Section \ref{sec_ion_emissivity} we discuss the intrinsic ionizing emissivities of galaxies and their dependence on the environment. Section \ref{sec_fesc} discusses our model where the ionizing escape fraction is proportional to the gas fraction ejected from the galaxy - here we show how $f_\mathrm{esc}$ depends on the mass and environment of a galaxy and its evolution during reionization. In Section \ref{sec_ionizing_budget} we bring our results from the previous two sections together and discuss which star-forming galaxies are the key drivers of reionization and how this result is affected by physical parameters such as the escape fraction dependency on galaxy properties and the radiative feedback strength. We conclude in Section \ref{sec_conclusions}. 
Throughout this paper we assume a $\Lambda$CDM Universe with cosmological parameter values of $\Omega_\Lambda=0.69$, $\Omega_m=0.31$, $\Omega_b=0.048$, $H_0=100h=67.8$km~s$^{-1}$Mpc$^{-1}$, $n_s=0.96$ and $\sigma_8=0.83$, and a Salpeter initial mass function \citep[IMF;][]{salpeter1955} between $0.1\msun$ to $100\msun$.

\section{The simulations}
\label{sec_sims}

In this paper, we use self-consistent, seminumerical simulations of galaxy evolution and reionization that were performed within the {\sc astraeus} framework. In this framework, we couple an N-body dark matter (DM) simulation, an enhanced version of the {\sc delphi} semi-analytic galaxy evolution model \citep{dayal2014} and a seminumerical reionization scheme {\sc cifog} \citep{hutter2018}. We refer the interested reader to \citet{hutter2021} for details of this framework. The key ingredients of {\sc astraeus}, which distinguish it from most other semi-analytical galaxy evolution codes, are ({\it i}) its self-consistent coupling to reionization with the radiative feedback strength of each galaxy depending on the time when its environment became reionized within the large-scale structure, ({\it ii}) the implementation of a wide range of plausible models for radiative feedback and the ionizing escape fraction, and ({\it iii}) its computational efficiency. Thanks to the short computing time, we can explore the impact of various physical processes on galaxies and reionization that would be hardly feasible with numerical simulations.

The underlying DM-only N-body simulation ({\sc very small multidark planck; vsmdpl}) is part of the {\sc multidark} simulation project\footnote{\url{https://www.cosmosim.org/}} and has been run using the {\sc gadget-2} Tree+PM (particle mesh) N-body code \citep{springel2005, klypin2016}. It has a box side length of $160h^{-1}$Mpc and follows the trajectories of $3840^3$ particles, with each particle having a DM mass of $m_\mathrm{DM}=6.2\times10^6h^{-1}\msun$. 
In total, 150 simulation snapshots have been stored from $z=25$ to $z=0$, with $74$ being saved between $z=25$ and $z=4.5$. For all snapshots, halos and subhalos down to $20$ particles have been identified using the {\sc rockstar} phase-space halo finder \citep{behroozi2013_rs}. The minimum resolved halo mass is $1.24\times10^8h^{-1}\msun$. Merger trees have been generated from the {\sc rockstar} catalogues by using {\sc consistent trees} \citep{behroozi2013_trees}. The vertical (sorted on a tree-branch-by-tree-branch basis within a tree) merger trees produced with {\sc consistent trees} have been resorted to local horizontal (sorted on a redshift-by-redshift-basis within a tree) merger trees using the {\sc cutnresort} module within the {\sc astraeus} pipeline. In addition, for all snapshots, the DM density fields have been produced by mapping the DM particles on to a $2048^3$ grid, which have been then re-sampled to the $512^3$ grid used as input files for the {\sc astraeus} code.

As described above, {\sc astraeus} couples the key physical processes of early galaxy formation {\it and} reionization \citep[for details see][]{hutter2021}. At each redshift-step, this includes processes on galactic scales, such as gas accretion, gas and stellar mass being brought in by mergers, star formation and associated SNII feedback as well as the large-scale reionization of the IGM and its associated radiative feedback on the gas content of early galaxies as now detailed. We assume each galaxy in a halo with mass $M_h$ to have an initial gas mass $M_\mathrm{g}^i(z)$ that corresponds to the cosmological baryon-to-DM ratio, $(\Omega_b/\Omega_m)~M_h$. In case a galaxy forms in an ionized region, $M_\mathrm{g}^i(z)$ is reduced to $M_\mathrm{g}^i(z) = f_\mathrm{g} (\Omega_b/\Omega_m)~M_h$, with $f_\mathrm{g}$ being the gas fraction that is not photo-evaporated by reionization. In subsequent time steps, a galaxy with $N_p$ progenitors can obtain gas through both smooth-accretion from the IGM as well as from mergers such that
\begin{eqnarray}
M_\mathrm{g}^\mathrm{acc}(z) & = & (\Omega_b/\Omega_m)\left[ M_h(z) - \sum_{p=1}^{N_p} M_{h,p}(z+\Delta z) \right]  \\ 
M_\mathrm{g}^\mathrm{mer}(z) & = & \sum_{p=1}^{N_p} M_\mathrm{g,p}(z+\Delta z),
\end{eqnarray}
where the first and second equations show the accreted and merged gas mass, respectively. Further, $M_{h,p}(z+\Delta z)$ and $M_\mathrm{g,p}(z+\Delta z)$ are the DM and final gas mass of the progenitors, respectively. However, we note that the resulting initial gas mass never exceeds the limit given by reionization feedback, $f_\mathrm{g} (\Omega_b/\Omega_m)~M_h$, and is given by
\begin{eqnarray}
    M_\mathrm{g}^i(z) &=& \min \left[ M_\mathrm{g}^\mathrm{mer}(z) + M_\mathrm{g}^\mathrm{acc}(z), f_\mathrm{g} \frac{\Omega_b}{\Omega_m} M_h(z) \right].
\end{eqnarray}

At each time step, a fraction of the initial gas mass, $f_\star^\mathrm{eff}=\min(f_\star, f_\star^\mathrm{ej})$, is converted into stellar mass, $M_\star^\mathrm{new}(z)=f_\star^\mathrm{eff} M_\mathrm{g}^i(z)$. Here, the effective star formation rate $f_\star^\mathrm{eff}$ can be thought of the fraction of gas mass that forms stars over a given timescale, and is given by the minimum that is required to eject all gas from the halo ($f_\star^\mathrm{ej}$) and an upper limit ($f_\star$). With $f_\star^\mathrm{eff}$ being linked to the gravitational potential of the halo, our model results in massive galaxies forming stars with a constant efficiency $f_\star$, while galaxies in low-mass halos are star formation efficiency limited through a combination of SNII and radiative feedback. While the factor $f_\mathrm{g}$ modifying the initial gas mass reservoir accounts for the effects of radiative feedback, the effective star formation rate accounts for the suppressed star formation in low-mass halos ($M_h\lesssim10^{9.5}\msun$) due to gas being heated and ejected from a galaxy by SNII explosions from star formation in the current and, since we account for mass-dependent stellar lifetimes \citet{padovani1993}, all previous time steps. We refer to this scheme as the delayed SN feedback scheme. In this delayed SN feedback limited regime, $f_\star^\mathrm{eff}$ is given by
\begin{eqnarray}
    f_\star^\mathrm{ej}(z) &=& \frac{v_c^2}{v_c^2 + f_w E_{51} \nu_z} \left[  1 - \frac{f_w E_{51} \sum_j \nu_j M_{\star,j}^\mathrm{new}(z_j)}{M_\mathrm{g}^i(z)~ v_c^2} \right]. 
    \label{eq_fej_delayedSN}
\end{eqnarray}
Here, $v_c$ is the rotational velocity of the halo, $E_{51}=10^{51}$erg~s$^{-1}$ is the energy produced by each SNII, $f_w$ is the fraction of SNII energy that couples to the gas and drives the winds, $M_{\star,j}^\mathrm{new}(z_j)$ is the newly formed stellar mass in time step $j$, and $\nu_j$ is the corresponding fraction that explodes as SNII in time step $z_j$ using our chosen IMF. From the star formation at the current and previous time steps, the gas mass ejected from the galaxy due to SNII feedback is calculated and the gas mass updated at each time step accordingly.

At each time step, we compute the spatial distribution of ionized regions, including the hydrogen ionization fractions and photoionization rates, with {\sc cifog}. For this purpose, we calculate the spectrum of each galaxy from its star formation history using the stellar population synthesis code {\sc starburst99} \citep{leitherer1999} assuming a Salpeter IMF and a metallicity of $0.05~Z_\odot$. The escaping ionizing emissivity is then derived as 
\begin{equation}
\dot{N}_\mathrm{ion}(z)=f_\mathrm{esc}\dot{Q}(z),
\end{equation}
where $\dot{Q}(z)$ is the intrinsic production rate of ionizing photons. From the ionizing emissivity (at each redshift step $\dot{N}_\mathrm{ion}$ of each galaxy is mapped onto a grid) and density grids, {\sc cifog} follows the approach outlined in \citet{furlanetto2004} and determines whether a grid cell is ionized or neutral by comparing the cumulative number of ionizing photons with the number of absorption events. Within the ionized regions {\sc cifog} derives the residual \HI fraction in each cell and the corresponding photoionization rate. Further details can be found in \citet{hutter2018}.
From the ionization and photoionization fields, we can determine on-the-fly whether the environment of a galaxy has been reionized and where applicable store the reionization redshift $z_\mathrm{reion}$. This is used to compute the corresponding radiative feedback strength in terms of a characteristic mass $M_c$ (defined as the mass at which a halo can maintain half of the gas mass compared to the cosmological baryon-to-DM ratio) and the corresponding gas mass not affected by reionization as $f_\mathrm{g}(\Omega_b/\Omega_m)M_h$ with $f_g=f(M_c, M_h)$.

Our framework has three free redshift-independent parameters: {\it (i)} the threshold star formation efficiency $f_\star$, {\it (ii)} the fraction of SN energy coupling to gas $f_w$, and {\it (iii)} the ionizing escape fraction $f_\mathrm{esc}$. These are tuned to reproduce observations, including the UV luminosity and stellar mass functions, and the specific star formation rate densities at $z=10-5$, constraints from Planck on the Thomson optical depth as well as constraints on the ionization state from Lyman-$\alpha$ emitters (LAEs), Gamma-ray bursts (GRBs) and quasars \citep[see Fig. 2-5][]{hutter2021}. Our simulation suite focuses on the effects of radiative feedback and contains different models for radiative feedback (or the corresponding characteristic mass $M_c$). However, since the time when the environment of a galaxy becomes ionized is critical for the strength of its radiative feedback, our simulations also comprise different models for the ionizing escape fraction. We briefly outline the characteristics of the different radiative feedback and ionizing escape fraction models used in this paper: \\
{\it (i) Photoionization} model: This model assumes a constant $f_\mathrm{esc}$ value for all galaxies and redshifts. $M_c$ is given by the fitting function that has been derived from the 1D radiation-hydrodynamical simulations in \citet{sobacchi2013a}. $M_c$ increases as the photoionization rate $\Gamma_\mathrm{HI}$ at $z_\mathrm{reion}$ and/or the difference between $z_\mathrm{reion}$ and the galaxy's current redshift $z$ rises. This model results in a weak to intermediate, time-delayed radiative feedback. \\
{\it (ii) Early Heating} model: This model assumes $f_\mathrm{esc}$ to scale with the gas fraction ejected from the galaxy such that $f_\mathrm{esc} = f_\mathrm{esc}^0 \min(1, f_\star^\mathrm{eff} / f_\star^\mathrm{ej})$ where $f_\mathrm{esc}^0=0.60$. $M_c$ equals the filtering mass $M_F$ that is described in \citet{gnedin1998b} and \citet{gnedin2000} and assumes that a region is heated up to $T=4\times10^4$~K upon reionization. This model results in a weak to intermediate time-delayed feedback. \\
{\it (iii) Strong Heating} model: This model assumes a constant $f_\mathrm{esc}$ value for all galaxies and redshifts. It also relates $M_c$ to the filtering mass $M_F$ but assumes $M_c=8 M_F$, as has been found in \citet{gnedin2000}. Ionized regions are also assumed to be heated up to $T=4\times10^4$~K. This model leads to a comparably strong, time-delayed radiative feedback, and represents our maximum time-delayed radiative feedback model. \\
{\it (iv) Jeans Mass} model: This model also assumes a constant $f_\mathrm{esc}$ value for all galaxies and redshifts. In contrast to all other radiative feedback models, $M_c$ is set to the Jeans mass $M_J(z)$ at virial over-density and $T=4\times10^4$~K {\it as soon as} the environment of a galaxy has been reionized. This model results in a strong, instantaneous radiative feedback, and represents our maximum instantaneous radiative feedback model.

We note that we investigate and discuss possible limitations of our model arising from the limited mass resolution of the {\sc vsmdpl} N-body simulation ($M_h\geq10^{8.26}\msun$) in detail in \citet{hutter2021} and Appendix \ref{app_reion_contribution_minihalos}.

\section{The environment dependent ionizing emissivity of galaxies}
\label{sec_ion_emissivity}

\begin{figure*}
\begin{minipage}{0.48\textwidth}
\includegraphics[width=\textwidth]{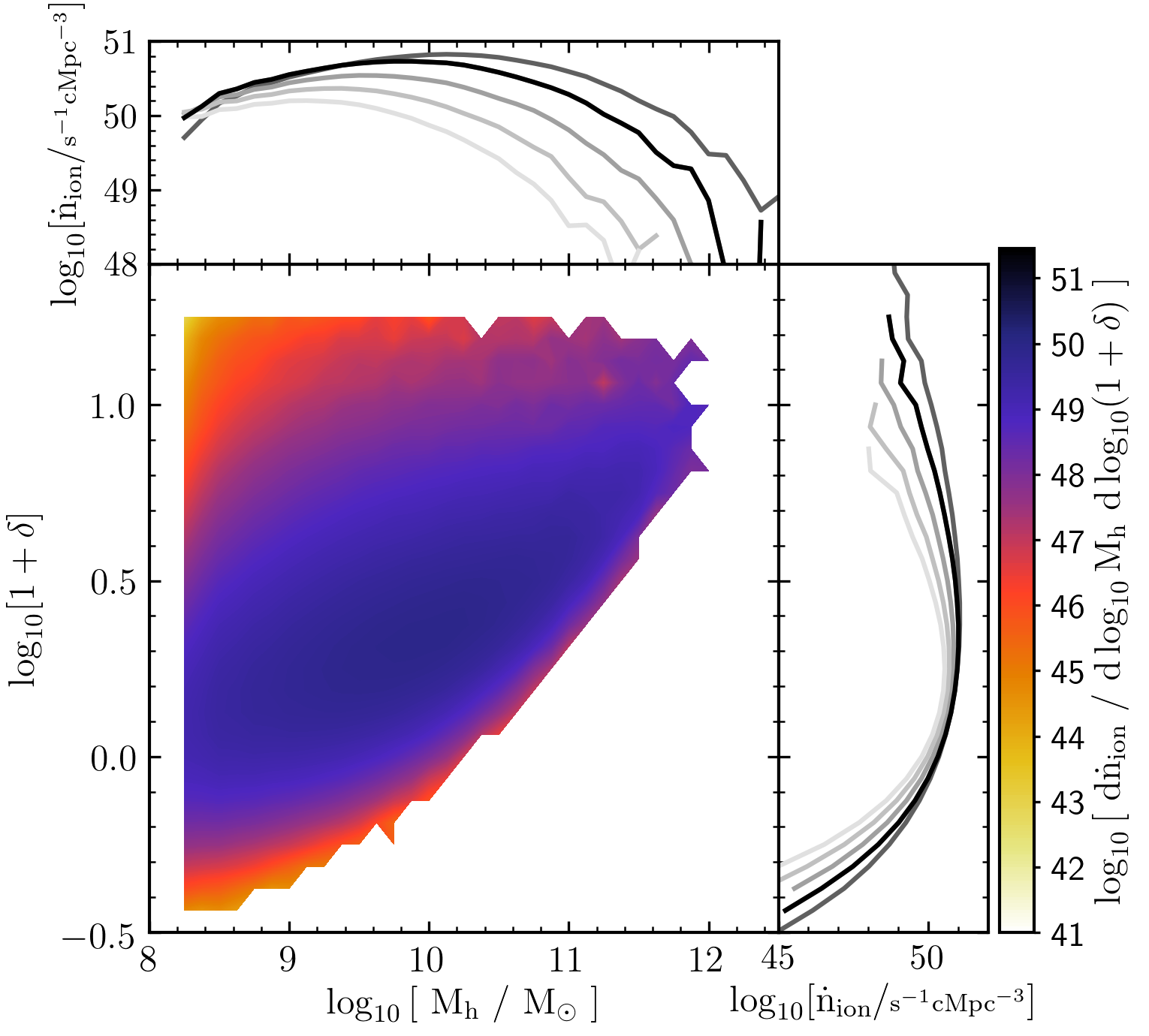}
\caption{The escaping ionizing emissivity density, $\dot n_\mathrm{ion}$, as a function of halo mass ($M_h$) and the over-density ($1+\delta$) smoothed over $2$~cMpc at $z=7$ for the {\it Photoionization} model is shown in the bottom left panel. The top left (bottom right) panel shows the 1D distributions of $\dot n_\mathrm{ion}$ as a function of $M_h$ ($1+\delta$) at $z=7$ (black line) and $z=6$, $8$, $9$, $10$ (increasingly lighter gray lines).}
\label{fig_Nion_dens_Mvir_SOBACCHI} 
\end{minipage}
\hspace{0.03\textwidth}
\begin{minipage}{0.48\textwidth}
\includegraphics[width=\textwidth]{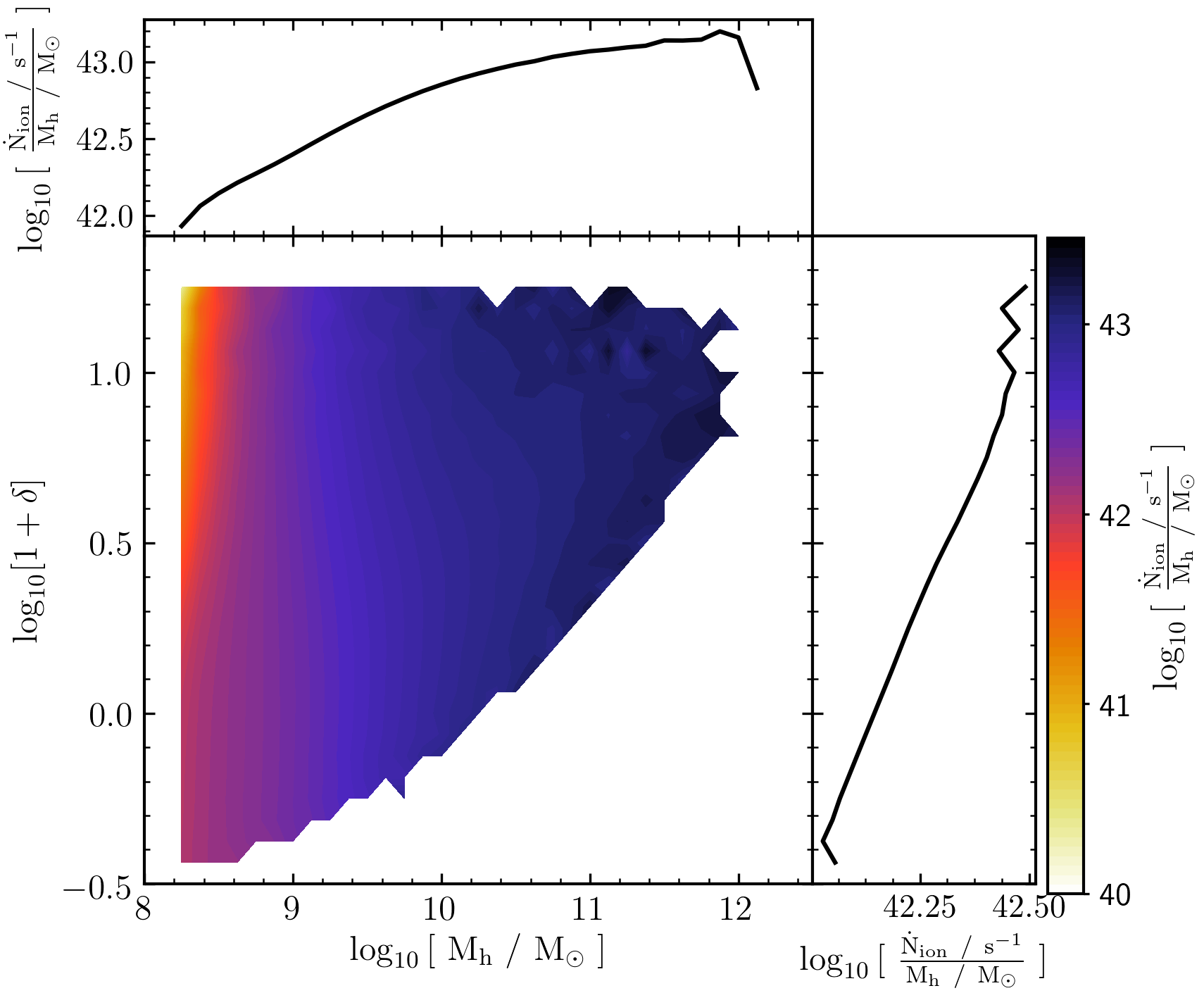}
\caption{The mean value of the ratio between the escaping ionizing emissivity ($\dot{N}_\mathrm{ion}$) and halo mass ($M_h$) as a function of $M_h$ and the over-density ($1+\delta$) smoothed over $2$~cMpc at $z=7$ for the {\it Photoionization} model is shown in the bottom left panel. The top left (bottom right) panels show the 1D distributions of $\dot{N}_\mathrm{ion}$ as a function of $M_h$ ($1+\delta$) at $z=7$ (black line).}
\label{fig_NionDIVMvir_dens_Mvir_SOBACCHI} 
\end{minipage}

\begin{minipage}{0.48\textwidth}
\includegraphics[width=\textwidth]{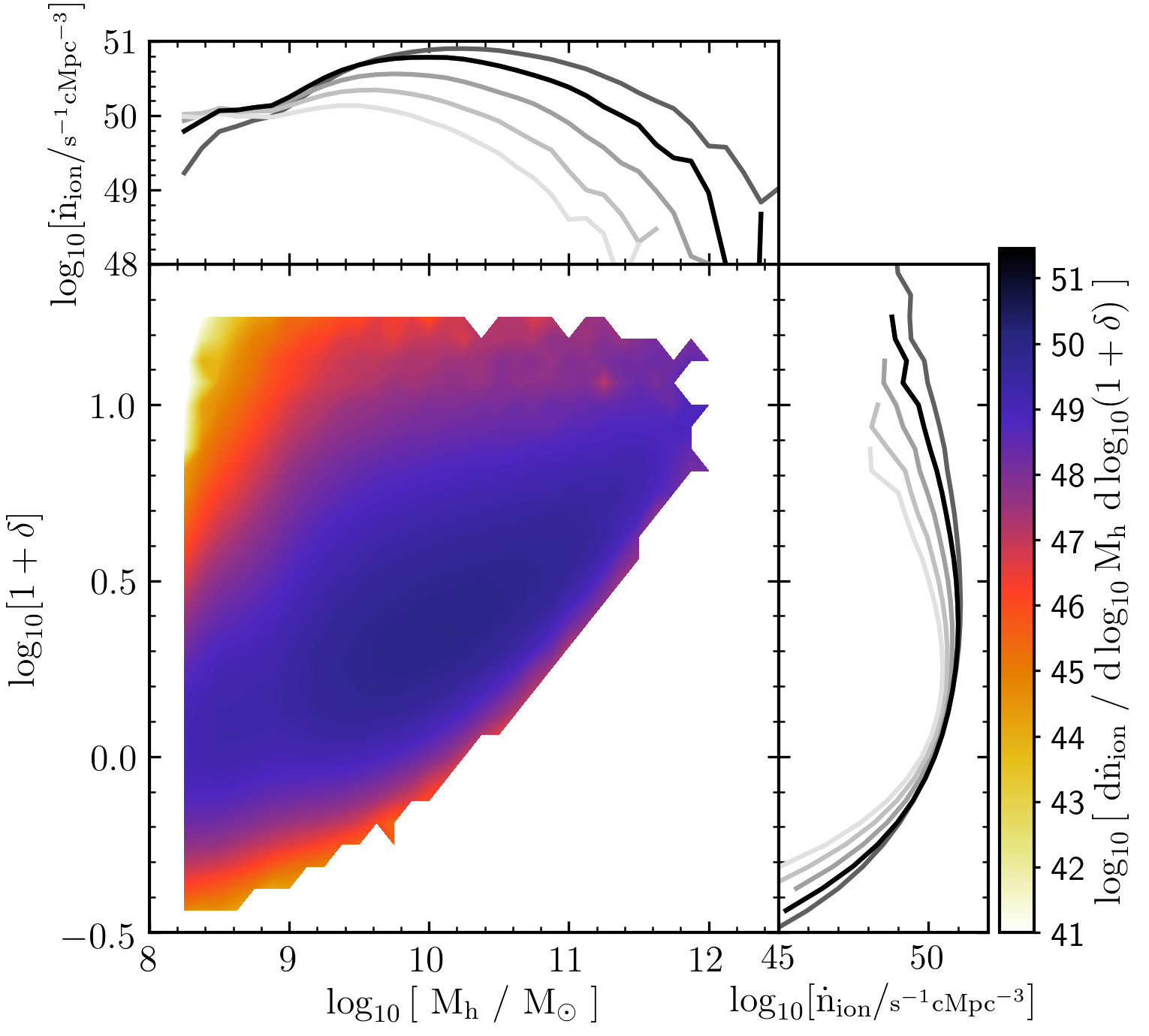}
\caption{The escaping ionizing photon emissivity density, $\dot n_\mathrm{ion}$, as a function of halo mass ($M_h$) and the over-density ($1+\delta$) smoothed over $2$~cMpc at $z=7$ for the {\it Jeans mass} model is shown in the bottom left panel. The top left (bottom right) panel shows the 1D distributions of $\dot n_\mathrm{ion}$ as a function of $M_h$ ($1+\delta$) at $z=7$ (black line) and $z=6$, $8$, $9$, $10$ (increasingly lighter gray lines).}
\label{fig_Nion_dens_Mvir_JEANSMASS}
\end{minipage}
\hspace{0.03\textwidth}
\begin{minipage}{0.48\textwidth}
\includegraphics[width=\textwidth]{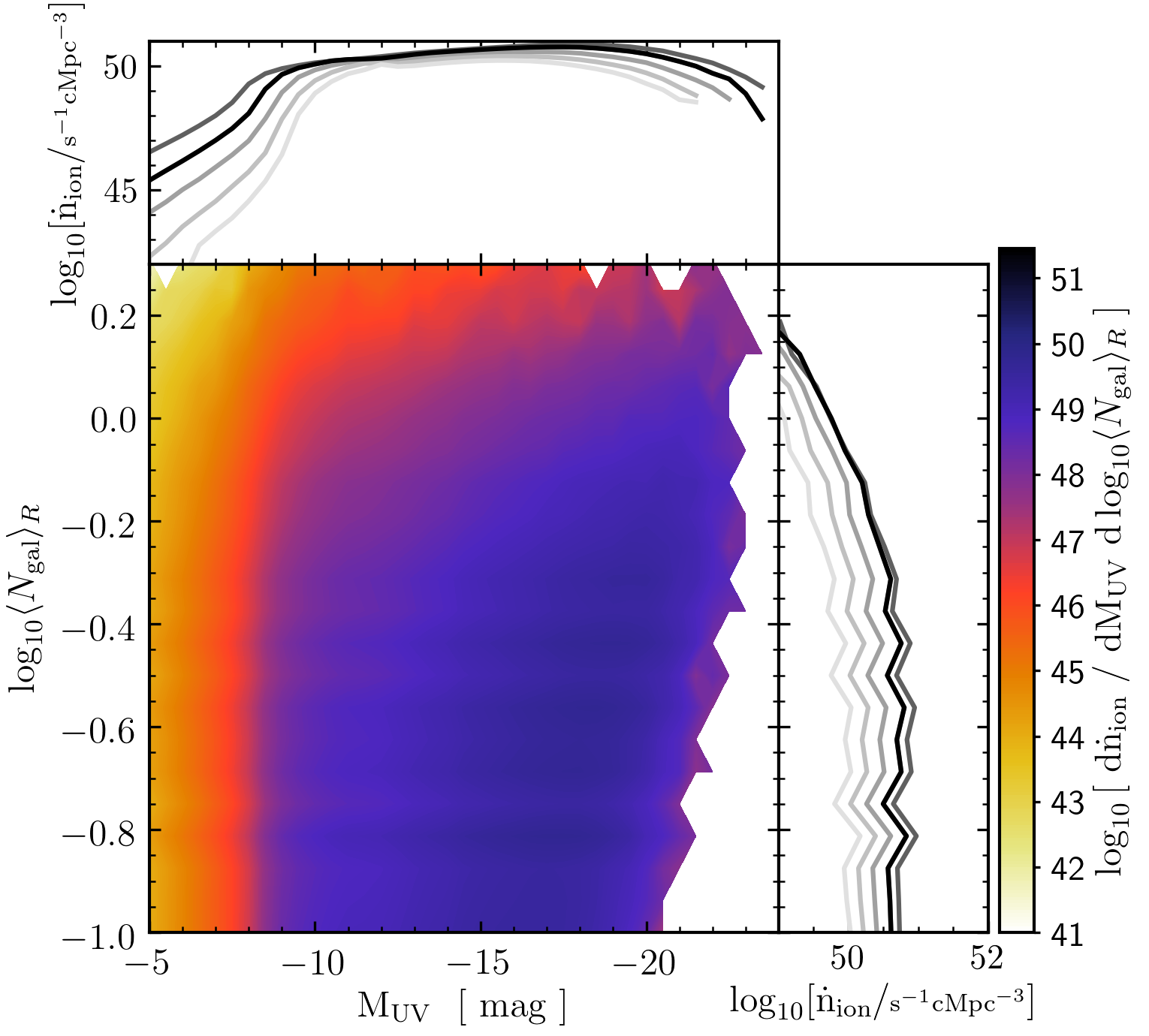}
\caption{The probability density distribution of the escaping ionizing emissivity density (${n}_\mathrm{ion}$) as a function of the UV luminosity of a galaxy ($M_\mathrm{UV}$) and the number of galaxies with $M_\mathrm{UV}\geq-14$ within a sphere of a $2$~cMpc diameter ($\langle N_\mathrm{gal}\rangle_R$) at $z=6$ is shown for the {\it Photoionization} model in the bottom left panel. The top left (bottom right) panels show the1D distributions of $\dot{n}_\mathrm{ion}$ as a function of $M_\mathrm{UV}$ (number of galaxies $\langle N_\mathrm{gal}\rangle_R$) at $z=7$ (black line) and $z=6$, $8$, $9$, $10$ (increasingly lighter gray lines).}
\label{fig_Nion_MUV_Ngal_SOBACCHI}
\end{minipage}
\end{figure*}

In this Section, we quantify which physical processes shape the contribution of ionizing photons from galaxies with different masses and environments. For this purpose, we analyse the ionizing emissivity as function of a key galactic property (halo mass and UV luminosity) and the environment of a galaxy (underlying density and number of neighbouring galaxies averaged over a sphere with a diameter of $2$~cMpc). Our smoothing scale of $2$~cMpc corresponds to the mean distance of galaxies with a halo mass of $M_h\simeq10^{9.5}\msun$ or a UV luminosity of $M_\mathrm{UV}\simeq-16$ at $z\simeq7$, which corresponds to the observational limit of the forthcoming James Webb Space Telescope (JWST)\footnote{We use the limiting magnitudes of the deep NIRcam/JWST imaging survey as part of JADES quoted in \citet{rieke2019, williams2018} as a guidance.}.
In the remainder of the paper, the over-density is expressed as $1+\delta = \rho / \langle\rho\rangle$. Here, $\rho$ is the average density within the $2$~Mpc sphere and $\langle\rho\rangle$ is the average density of the Universe at the respective redshift.

\subsection{Ionizing emissivity as a function of halo mass \& density}
\label{subsec_ion_emissivity_Mh_density}

In this section, we discuss the distribution of the escaping ionizing emissivity density which is described as 
\begin{equation}
\dot n_\mathrm{ion}=\frac{\sum_{i=1}^{N_\mathrm{gal}} \dot{N}_{\mathrm{ion},i}}{V}=\frac{\sum_{i=1}^{N_\mathrm{gal}} f_{\mathrm{esc},i} \dot{Q}_i}{V} \ \ [\mathrm{s}^{-1}\mathrm{Mpc}^{-3}],
\end{equation} 
where $ \dot{N}_{\mathrm{ion},i}$ is the escaping ionizing emissivity from the $i^\mathrm{th}$ galaxy and $N_\mathrm{gal}$ the number of galaxies for a selected sample (e.g. a halo mass - density bin), and $V$ the volume of our simulation box. We start by discussing our results at redshift $z=7$ (that roughly marks the mid-point of reionization 
for all our models) for the {\it Photoionization model} as shown in Fig. \ref{fig_Nion_dens_Mvir_SOBACCHI}. Here, we show the escaping ionizing emissivity density distribution as a function of the halo mass ($M_h$) and over-density $\delta$ (bottom left panel) in addition to the respective 1D distributions as a function of halo mass (top left panel) and over-density (bottom right panel). 

From the 1D distributions, we see that most escaping ionizing photons come from intermediate mass galaxies in slightly over-dense regions, with their distributions peaking at $\log_{10}(1+\delta)\simeq0.2-0.5$ and $M_h\simeq10^{10}\msun$, respectively. The reason that intermediate mass galaxies dominate the ionizing emissivity density is due to a combination of their being numerous enough and having a deep enough gravitational potential so that their gas content and star formation rates are hardly affected by either SNII- or radiative-feedback. While less massive galaxies have a far higher abundance, their star formation rates and, hence, ionizing emissivities are largely suppressed by SNII- and radiative-feedback; although being unaffected by either type of feedback, more massive galaxies are just far rarer.

The bottom left panel in Fig. \ref{fig_Nion_dens_Mvir_SOBACCHI} allows us to draw a more differentiated picture on the interplay between galaxy properties (halo mass) and the environment (density). We can see that the majority of ionizing photons come from galaxies whose halo masses correspond to the underlying density and follow the ``average" distribution (as shown by the 50\% line) in Fig. \ref{fig_Ngal_dens_Mvir_SOBACCHI}. We refer to this as the ``halo mass - underlying density relation" i.e. the escaping ionizing emissivity is dominated by low-mass galaxies in less dense regions, intermediate mass galaxies in intermediate-density regions, and high mass galaxies in very over-dense regions; we note that this trend persists to higher redshifts.
The reason for this trend is twofold as now detailed: firstly and above all, the trend follows the number density of galaxies closely (see Fig. \ref{fig_Ngal_dens_Mvir_SOBACCHI} in Appendix \ref{app_ngal}), i.e. there are up to $2-4$ orders of magnitude less low-mass galaxies ($M_h\lesssim10^{9.5}\msun$) in over-dense than in average ($\log_{10}(1+\delta)\simeq0$) and intermediate ($\log_{10}(1+\delta)\simeq0.5$) density regions. Similarly, intermediate mass galaxies ($M_h\simeq10^{9.5-11}\msun$) are far more abundant in intermediate-density than under-dense regions, while high mass galaxies ($M_h\gtrsim 10^{11}\msun$) only reside in over-dense regions.

Secondly, for low-mass galaxies, the trend is enhanced by the fact that reionization feedback is more efficient in suppressing star formation in low-mass galaxies in over-dense regions as compared to under-dense ones. In our framework, over-dense regions are ionized earlier than under-dense regions; this results in radiative feedback being more effective in suppressing star formation in over-dense regions due to the higher characteristic mass $M_c$ \citep[see][]{hutter2021}. 
This effect can be seen in Fig. \ref{fig_NionDIVMvir_dens_Mvir_SOBACCHI}, where we show the average ratio between the escaping ionizing emissivity and halo mass as a function of halo mass and over-density. The ratio $\dot N_\mathrm{ion}/M_h$ decreases not only towards lower halo masses due to SN feedback increasingly suppressing star formation, but also towards over-dense regions due to a stronger radiative feedback: it drops from $\sim10^{42.2}$s$^{-1}\msun^{-1}$ at a density of $\log_{10}(1+\delta)\simeq 0$ to $\sim10^{41}$s$^{-1}\msun^{-1}$ at $\log_{10}(1+\delta)\gtrsim1$ for galaxies with $M_h<10^{8.4}\msun$.

As for galaxies that do not lie on the halo mass - underlying density relation (i.e.  low- and intermediate-mass galaxies in over-dense and under-dense regions, respectively), the ionizing emissivity is also dependent on their gas accretion histories. Galaxies in over-dense regions have lower dark matter and gas mass accretion rates but higher ionizing emissivities than galaxies in less dense regions as shown in Fig. \ref{fig_NionDIVMvir_dens_Mvir_SOBACCHI} for $M_h\simeq10^{9-11}\msun$ halos. The reason for the reduced gas accretion in over-dense regions is probably due to the fact that the unbound gas feels more gravitational attraction towards neighbouring massive galaxies than to low-mass galaxies. However, this trend is negligible when compared to those of galaxy number densities and radiative feedback and hardly affects the distribution of the ionizing emissivity as a function of halo mass and the underlying density, which is why we defer an in-depth analysis to a future work (Legrand et al., in prep.).

We now discuss the role of radiative feedback in suppressing the contribution of low-mass galaxies in dense regions for the different radiative feedback models used. As the strength of radiative feedback rises, going from the {\it Photoionization} model (c.f. Fig. \ref{fig_Nion_dens_Mvir_SOBACCHI}) to the {\it Jeans Mass} model (c.f. Fig. \ref{fig_Nion_dens_Mvir_JEANSMASS}), star formation and hence the ionizing emissivity in low-mass halos are increasingly suppressed over time. While low-mass halos (with $M_h \lsim 10^9 \msun$) contribute $\dot n_\mathrm{ion} = 10^{44-47} {\rm s^{-2}}\msun^{-1}$cMpc$^{-3}$ in the {\it Photoionization} model in the most over-dense regions ($\log_{10}(1+\delta)\gsim 1$), they contribute $\dot n_\mathrm{ion}\simeq10^{41-46}{\rm s^{-2}}\msun^{-1}$cMpc$^{-3}$ in the {\it Jeans Mass} model for the same halo mass and density values.

Next we discuss how the contribution from galaxies with different halo masses and underlying densities changes throughout reionization (see Figs. \ref{fig_Nion_dens_Mvir_SOBACCHI} and \ref{fig_Nion_dens_Mvir_JEANSMASS}). As cosmic time proceeds, the density of the Universe decreases, while the density contrast increases, i.e. over-dense regions become more over-dense (by a factor of $\sim3.5$ from $z=10$ to $6$) while under-dense regions become even less dense (by a factor of $\sim 0.7$ from $z=10$ to $6$). This evolution results in galaxies becoming continually more massive in over-dense regions with more low-mass galaxies forming in the newly collapsing over-density peaks. 

Firstly, this evolution is reflected in the distribution of the ionizing emissivity density as a function of the underlying density field, which extends to more over- and under-dense regions as redshift decreases (light to dark gray lines in the bottom right panels), ranging from low-mass (massive) galaxies with $M_h=10^9\msun$ ($M_h=10^{11}\msun$) being located in $\log_{10}(1+\delta)\sim-0.2$ to $0.9$ ($0.4$ to $0.7$) at $z=10$ to $\log_{10}(1+\delta)\sim-0.4$ to $1.5$ ($0.2$ to $1.4$) at $z=6$. Furthermore, while $\sim90$\% of massive galaxies ($M_h\sim10^{11}\msun$) are located in over-dense regions with $\log_{10}(1+\delta)\gtrsim0.5$ ($0.45$) at $z=10$ ($6$) and populate more over-dense regions as redshift decreases, just $10$\% of low-mass galaxies ($M_h\sim10^9\msun$) populate these over-dense regions with $\log_{10}(1+\delta)\gtrsim0.4$ ($0.5$) at $z=10$ ($6$) (c.f. Fig. \ref{fig_Ngal_dens_Mvir_SOBACCHI} for $z=6$).

Secondly, the distribution of the ionizing emissivity density as a function of halo mass extends to higher halo masses with decreasing redshift (from light to dark gray lines in the top left panels), given that an increase in galaxy mass results in an increase of stellar mass and the number of ionizing photons. However, despite more low-mass halos forming with decreasing redshift, we find that their contribution to the overall ionizing emissivity drops and the halo mass with the highest output of ionizing photons, $M_h^\mathrm{n_{ion,p}}$, shifts to larger halo masses from $M_h\simeq10^{9.2}\msun$ at $z=10$ to $M_h\simeq10^{10.2}\msun$ at $z=6$, which is in agreement with the intrinsic emissivity results from the {\sc CoDa ii} simulation in \citet{lewis2020}. This decrease in the relative contribution of low-mass galaxies to the ionizing budget is caused by the increasing SN and radiative feedback strength with time that decreases their gas mass; this also propagates through to the gas content of their successor halos at later redshifts. On the one hand, as the density of the Universe decreases, the gravitational potentials of halos with the same halo mass $M_h$ flatten, allowing larger fractions of gas (about a factor of $1.6$ higher at $z=6$ than at $z=10$) being ejected from galaxies as the same number of SNII explode; i.e. SN feedback affects increasingly massive galaxies with cosmic time. On the other hand, radiative feedback becomes stronger (increase in its characteristic mass $M_c$) the longer a galaxy's environment has been reionized, since the gas in that galaxy has had more time to adapt its density to its increased temperature upon reionization. It is expected that as we increase the radiative feedback strength (by adopting e.g. a higher temperature the gas in ionized regions is heated up to or assuming no time-delay in the gas density in the galaxy adjusting to its higher temperature), i.e. going from the {\it Photoionization} to the {\it Strong Heating} to the {\it Jeans Mass} models, the contribution of low-mass galaxies to the ionizing budget will drop, as can be seen when comparing the top left panels of Figs. \ref{fig_Nion_dens_Mvir_SOBACCHI} and \ref{fig_Nion_dens_Mvir_JEANSMASS}. 

\subsection{Ionizing emissivity as a function of luminosity \& number of galaxies} 
\label{subsec_ion_emissivity_MUV_Ngal}

In this Section, we discuss the ionizing emissivity as a function of two observable quantities, the absolute UV magnitude ($M_\mathrm{UV}$) and the 
number of galaxies in a sphere with a diameter of $2$~cMpc centered around the respective galaxy ($\langle N_\mathrm{gal} \rangle_R$); these are effectively proxies for the halo mass and the underlying density field, respectively. We start by discussing the ionizing emissivity density as a function of $M_\mathrm{UV}$ and $\langle N_\mathrm{gal}\rangle_R$ for the {\it Photoionization} model as shown in Fig. \ref{fig_Nion_MUV_Ngal_SOBACCHI}. As expected, we find very similar trends as those discussed above in Sec. \ref{subsec_ion_emissivity_Mh_density}. 

Firstly, from the ionizing emissivity density distribution as a function of $M_\mathrm{UV}$ (see top left panel in Fig. \ref{fig_Nion_MUV_Ngal_SOBACCHI}), we see that most ionizing photons are provided by intermediate bright galaxies ($M_\mathrm{UV}\simeq-18$) at $z=6-10$. However, with the curve being rather shallow between $M_\mathrm{UV}\simeq-14$ to $-20$, the contribution of both fainter and brighter galaxies remains significant. This is because while the low number density of bright galaxies ($n\simeq10^{-3.5}$cMpc$^{-3}$ for $M_\mathrm{UV}=-20$ at $z=7$) is compensated by their steady star formation (that is hardly affected by SN and radiative feedback), the reduced ionizing emissivity of star formation suppressed/feedback-limited fainter galaxies is balanced by their higher number densities ($n\simeq10^{-1}$cMpc$^{-3}$ for $M_\mathrm{UV}=-14$ at $z=7$).
The prevalent location of this galaxy population in the large-scale structure given by $\langle N_\mathrm{gal} \rangle_R$ is seen in the bottom left panel in Fig. \ref{fig_Nion_MUV_Ngal_SOBACCHI}.
For our constant $f_\mathrm{esc}$ scenarios, the galaxies in the $M_\mathrm{UV}$-$\langle N_\mathrm{gal} \rangle_R$ plane contributing most to the overall ionizing emissivity echo the ``halo mass - underlying density relation" explained in the previous Section: qualitatively, brighter galaxies contributing most to the overall ionizing emissivity lie in regions with a higher number of neighbours (from $\langle N_\mathrm{gal}\rangle_R\sim0.4$ for $M_\mathrm{UV}\simeq-19$ to $\langle N_\mathrm{gal} \rangle_R\sim1$ for $M_\mathrm{UV}\simeq-21$, corresponding to denser regions), while fainter galaxies contributing most to the overall ionizing emissivity lie in regions with a lower number of neighbours (from $\langle N_\mathrm{gal}\rangle_R\sim0.2$ for $M_\mathrm{UV}\simeq-17$ to $\langle N_\mathrm{gal}\rangle_R\sim0.1$ for $M_\mathrm{UV}\simeq-15$, corresponding to less dense regions).

Secondly, as can be seen from the bottom left panel of the same plot, the contribution of faint galaxies ($M_\mathrm{UV}\gtrsim-15$) to the total ionizing emissivity decreases as the number of galaxies in their environment, $\langle N_\mathrm{gal} \rangle_R$, (and hence their underlying density) increase. Similar to the findings for Fig. \ref{fig_Nion_dens_Mvir_SOBACCHI}, this trend is mainly due to a combination of a rapid decline in the number of low-mass galaxies in increasingly denser (higher $\langle N_\mathrm{gal} \rangle_R$ values) regions and due to radiative feedback having a stronger effect on low-mass galaxies in over-dense regions (since their environments are reionized earlier).

Thirdly, from the top left panel of the same figure we see that the contribution of bright galaxies increases as the redshift decreases and the Universe becomes  increasingly more ionized. This trend is due to a combination of these galaxies continuously growing in dark mass (and hence the ionizing emissivity) as well as the decreasing contribution from low-mass galaxies.

We briefly note that, despite these similarities, the trend of the ionizing emissivity density as a function of halo mass at the low-mass end (top left panel in Fig. \ref{fig_Nion_dens_Mvir_SOBACCHI}) differs from that as a function of UV magnitude at the low-luminosity end (top left panel in Fig. \ref{fig_Nion_MUV_Ngal_SOBACCHI}). This is because, while the contribution of low-mass galaxies increases marginally with decreasing redshift, the contribution of faint galaxies increases noticeably. This can be explained as follows: the increasing impact of radiative feedback with decreasing redshift (due to more low-mass halos being located in ionized regions and the increase of the radiative feedback characteristic mass with time), coupled with SNII feedback, leads to a continual reduction in the star formation rates of low-mass galaxies, i.e. for a given halo mass, galaxies have a lower UV luminosity and ionizing emissivity with decreasing redshift (due to a larger fraction of older stellar populations). In other words, the range of star formation rate values for a given halo mass broadens and this broadening shifts to higher halo masses as redshift decreases. Hence, as cosmic time passes, an increasing fraction of increasingly more massive halos will show the same UV luminosity as the low-mass halos at earlier times, and contribute to the ionizing emissivity originating from faint galaxies.

\subsection*{}
In summary, for constant $f_\mathrm{esc}$ scenarios, galaxies with intermediate masses ($M_h\simeq10^{9-11}\msun$) and absolute magnitudes ($M_\mathrm{UV}\simeq-15$ to $-20$) in average- to intermediate-density regions ($\log_{10}(1+\delta) \simeq 0-0.8$) drive reionization. This result is in good agreement with the masses and UV luminosities of galaxies contributing most to reionization that were found in \citet{yung2020}.
Despite being more abundant, lower-mass/fainter galaxies are very inefficient in forming stars due to a combination of SN and radiative feedback processes while higher-mass/brighter galaxies are just too rare despite their higher star formation rates. Hence, affecting only lower mass galaxies, radiative feedback has a negligible effect on shaping the galaxies driving reionization. Interestingly, however, radiative feedback has an environment-dependent effect on the ionizing emissivity of low-mass halos: being reionized earlier, low-mass halos in over-dense regions show a larger suppression of the ionizing emissivity. For our {\it Photoionization} model, radiative feedback decreases the ionizing emissivity of the lowest-mass galaxies ($M_h\lesssim10^{8.5}\msun$) by an order of magnitude when going from average ($\log_{10}(1+\delta)\simeq0$) to over-dense regions ($\log_{10}(1+\delta)\simeq1$).

\section{Linking the ionizing escape fraction to gas ejection}
\label{sec_fesc}

\begin{figure*}
\begin{minipage}{0.48\textwidth}
\includegraphics[width=\textwidth]{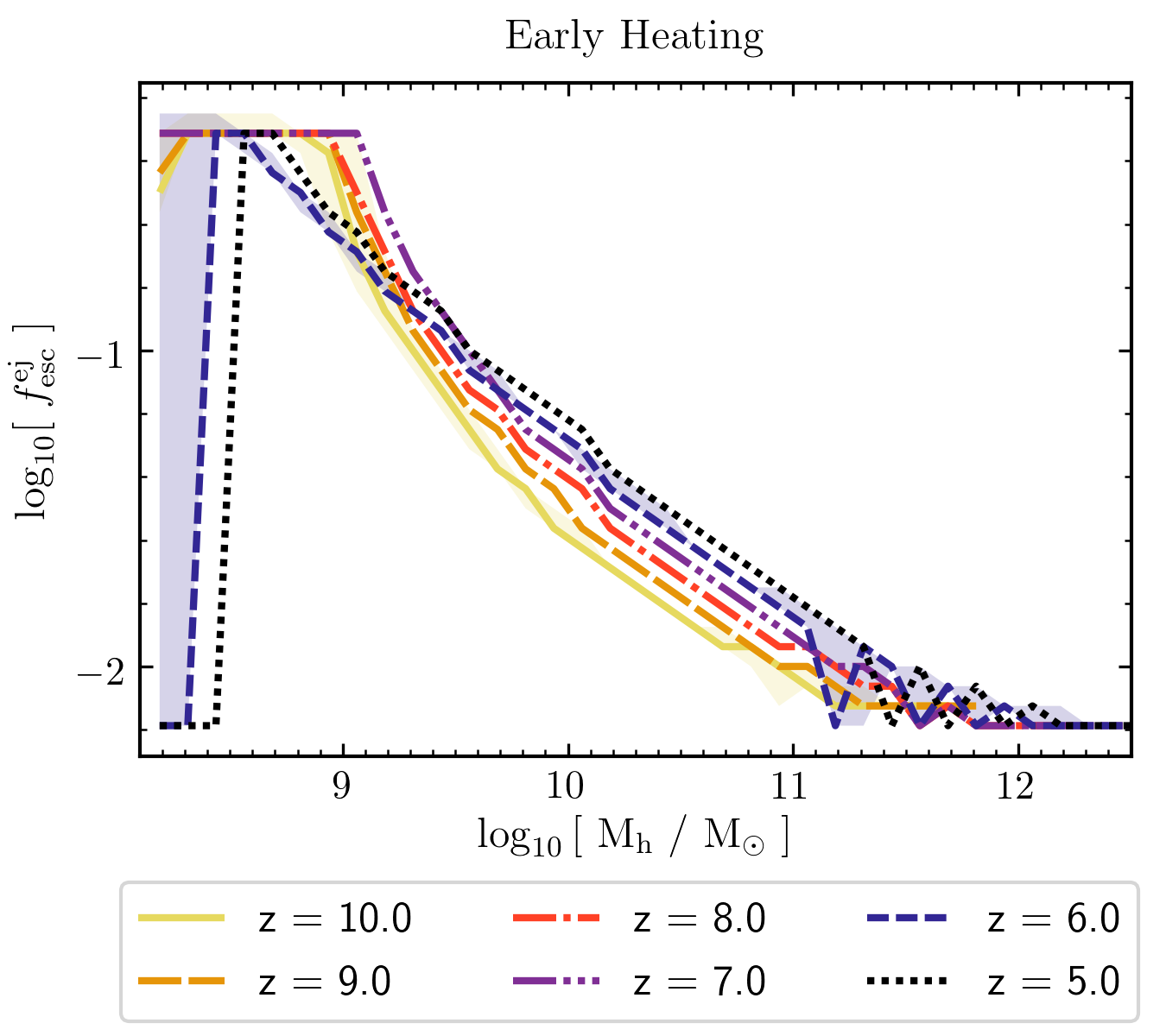}
\caption{The ionizing escape fraction as a function of halo mass at $z=10$, $9$, $8$, $7$, $6$, and $5$, as marked, for the {\it Early Heating} model. The coloured lines show the median of the distributions, while the transparent shaded coloured regions mark the $1-\sigma$ limits of the $f_\mathrm{esc}^\mathrm{ej}$ values at $z=10$ and $z=6$.}
\label{fig_fesc_Mvir_FESCTEMPEVOL} 
\end{minipage}
\hspace{0.03\textwidth}
\begin{minipage}{0.48\textwidth}
\includegraphics[width=\textwidth]{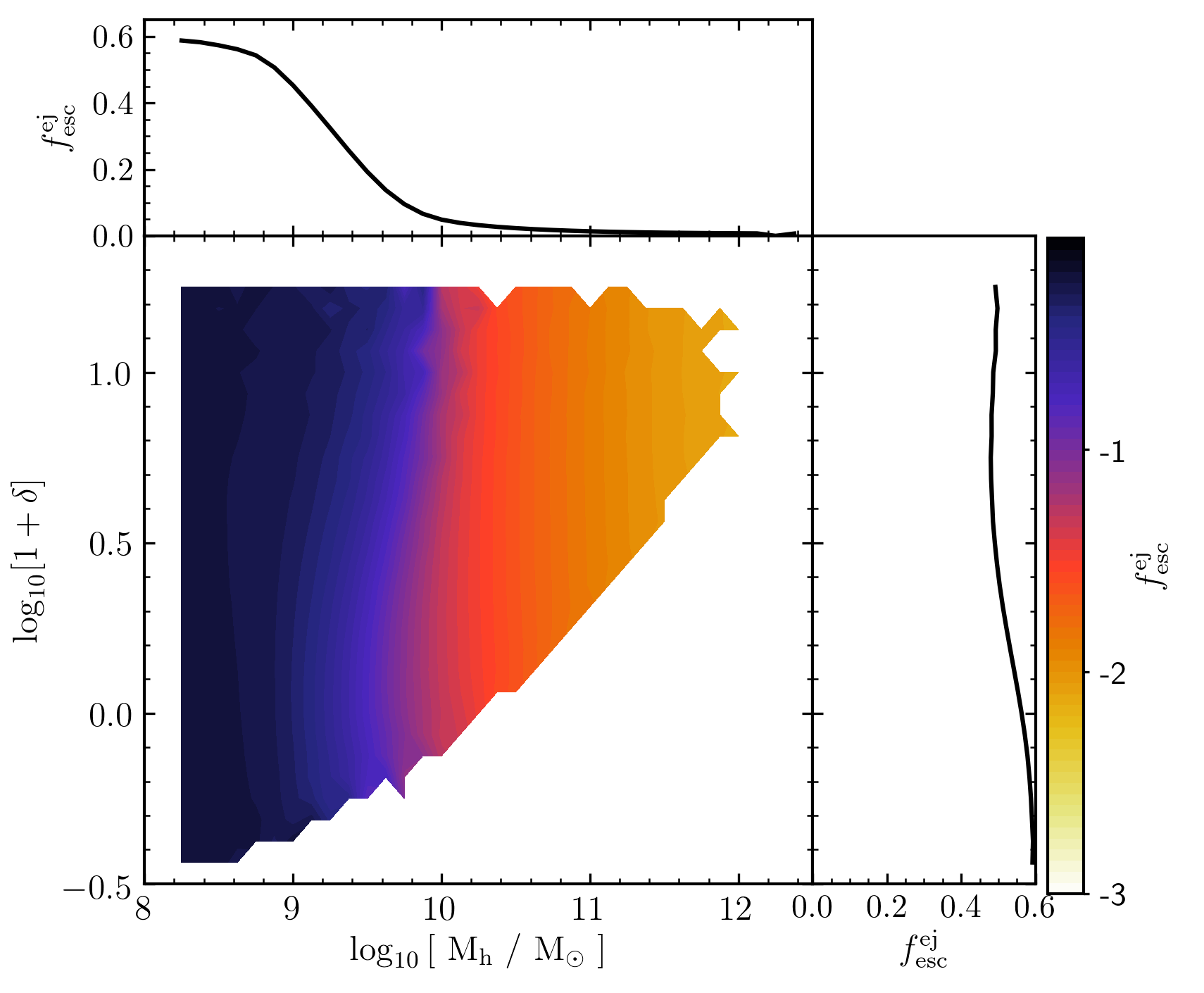}
\caption{The ionizing escape fraction ($f_\mathrm{esc}^\mathrm{ej}$) as a function of halo mass and density at $z=7$ for the {\it Early Heating} model. In the bottom left panel we show the mean value of $f_\mathrm{esc}^{ej}$ colour-coded as a function of halo mass and density. The top left and bottom right panel show the 1D distribution functions of $f_\mathrm{esc}^\mathrm{ej}$ as a function of the halo mass and underlying density, respectively.}
\label{fig_fesc_dens_Mvir_FESCTEMPEVOL} 
\end{minipage}

\begin{minipage}{0.48\textwidth}
\includegraphics[width=\textwidth]{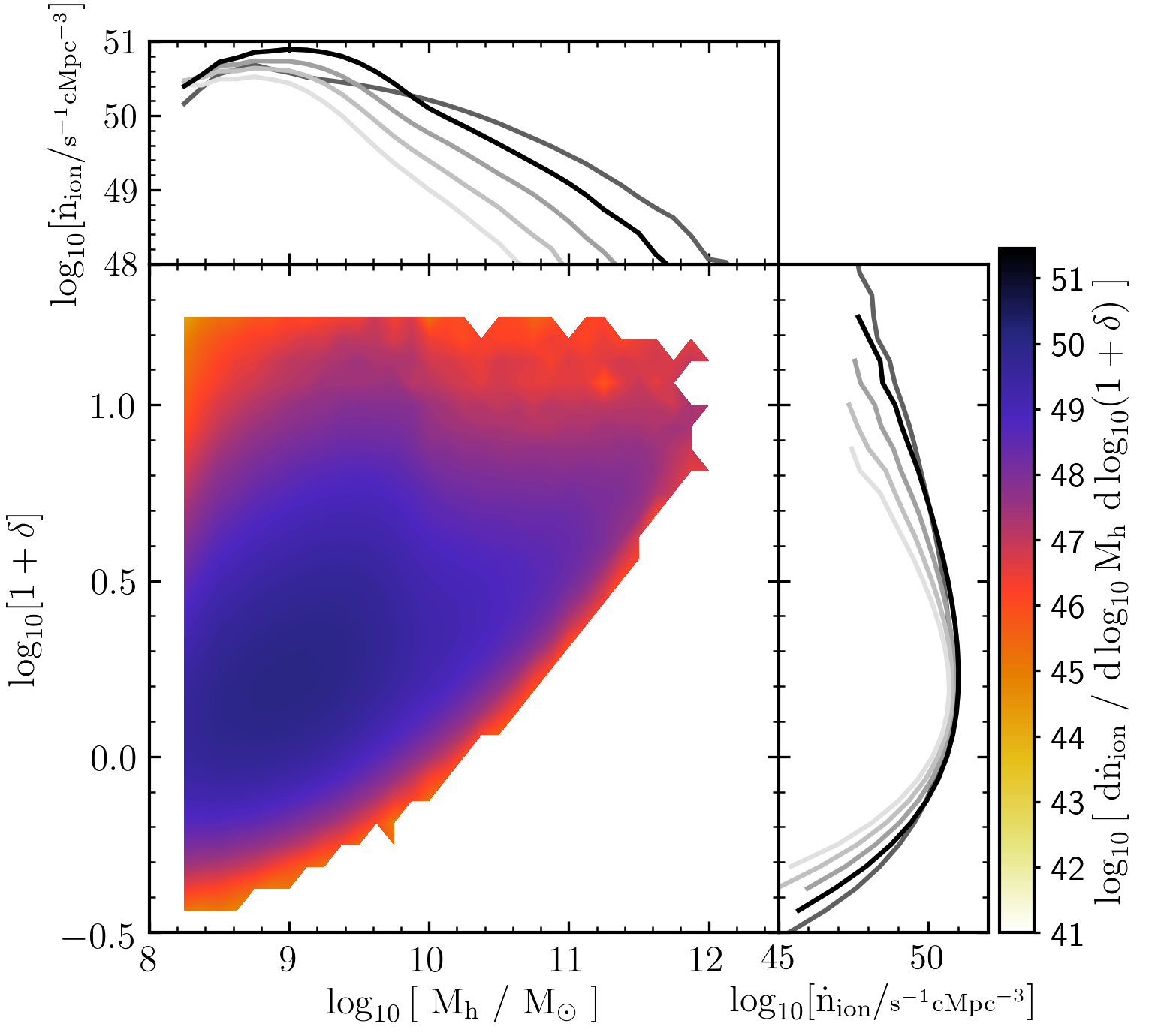}
\caption{The escaping ionizing emissivity density as a function of halo mass and the over-density smoothed over $2$~cMpc at $z=7$ is shown for the {\it Early Heating} model in the bottom left panel. The top left (bottom right) panels show the 1D distributions of the escaping ionizing emissivity density as a function of halo mass ($1+\delta$) at $z=7$ (black line) and $z=6$, $8$, $9$, $10$ (increasingly lighter gray lines).}
\label{fig_Nion_dens_Mvir_FESCTEMPEVOL} 
\end{minipage}
\hspace{0.03\textwidth}
\begin{minipage}{0.48\textwidth}
\includegraphics[width=\textwidth]{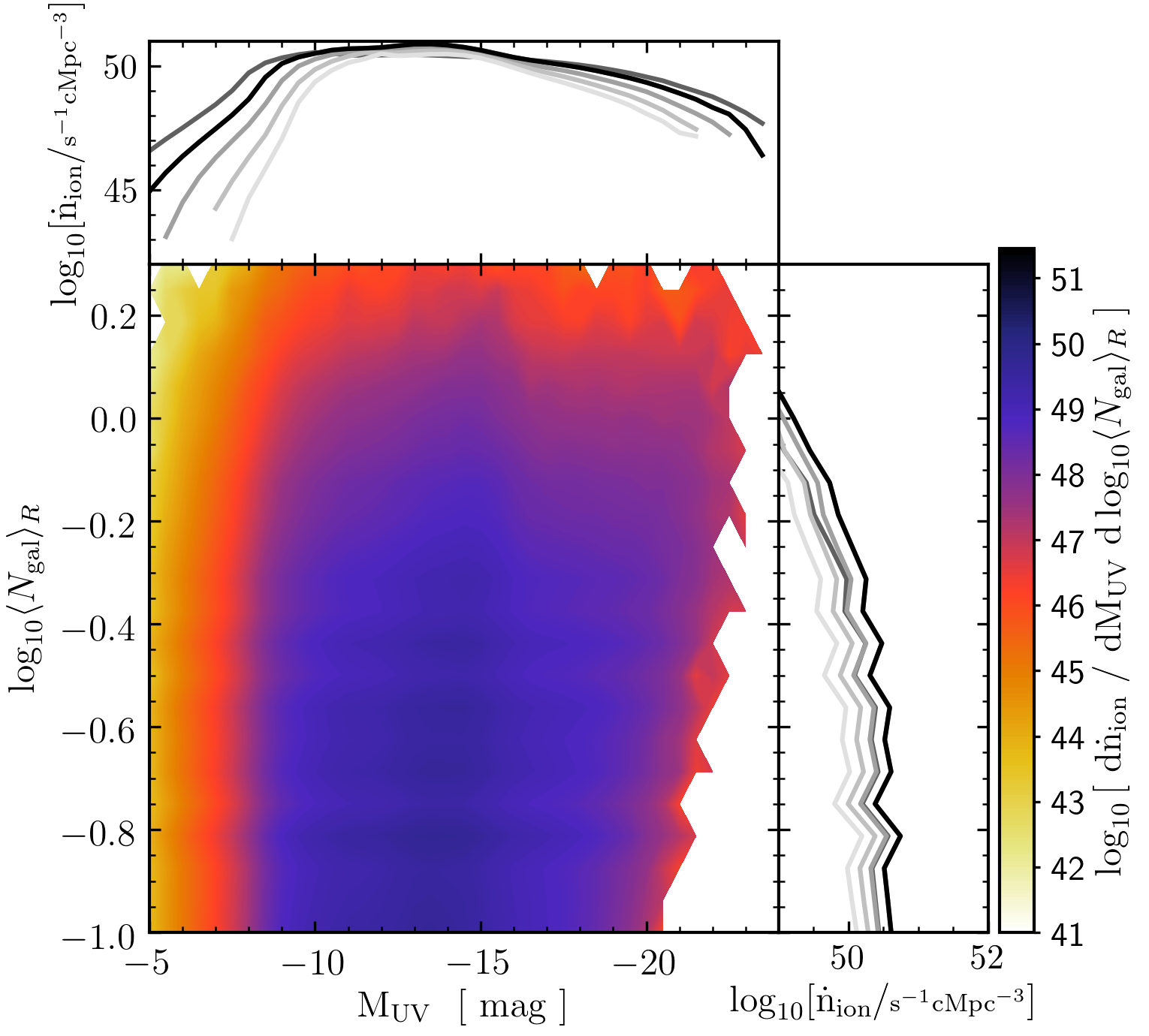}
\caption{The escaping ionizing emissivity density as a function of the UV luminosity and the number of galaxies with $M_\mathrm{UV}\geq-14$ within a sphere of a $2$~cMpc diameter at $z=7$ is shown for the {\it Early Heating} model in the bottom left panel. The top left (bottom right) panels show the 1D distributions of this quantity as a function of $M_\mathrm{UV}$ ($\langle N_\mathrm{gal}\rangle_R$) at $z=7$ (black line) and $z=6$, $8$, $9$, $10$ (increasingly lighter gray lines).}
\label{fig_Nion_MUV_Ngal_FESCTEMPEVOL} 
\end{minipage}
\end{figure*}

In this Section, we explore how a mass-dependent $f_\mathrm{esc}$ can affect the galaxy population that drives reionization. Multiple radiation hydrodynamical simulations have shown that, as a result of SN explosions being more effective in ejecting gas in lower gravitational potentials, low-mass galaxies typically exhibit more low-density tunnels than high-mass galaxies. This results in a $f_\mathrm{esc}$ value that increases with decreasing halo mass \citep{paardekooper2015, Kimm2014, Kimm2017}. Based on this finding, we link $f_\mathrm{esc}$ to the gas fraction that is ejected from the galaxy, $f_\mathrm{esc}^\mathrm{ej}=f_\mathrm{esc}^0 \min(1, f_\star^\mathrm{eff}/f_\star^\mathrm{ej})$ - this is our {\it Early Heating} model.

In Fig. \ref{fig_fesc_Mvir_FESCTEMPEVOL} we show the corresponding evolution of $f_\mathrm{esc}^\mathrm{ej}$ as a function of halo mass $M_h$ for $z \sim 6-10$.
From this figure, we can see that the ionizing escape fraction $f_\mathrm{esc}^\mathrm{ej}$ decreases with increasing halo mass. This behaviour is expected, since we have linked $f_\mathrm{esc}^\mathrm{ej}$ to the fraction of gas that is ejected from the galaxy: as a galaxy's gravitational potential deepens, a smaller gas fraction is ejected. In terms of the redshift evolution of $f_\mathrm{esc}^\mathrm{ej}$, we find that for galaxies with $M_h\gtrsim10^{9.7}\msun$, at a given halo mass, the median ionizing escape fraction increases with decreasing redshift. This is essentially because halos of a given mass are spatially more extended (i.e. have a shallower potential) with decreasing redshift which allows a larger gas fraction to be ejected from the galaxy, leading to an increase in $f_\mathrm{esc}^\mathrm{ej}$. While we would expect to see this trend across all times and halo masses (and indeed we can see it for $z=10-7$ and $z=6-5$), we do not find it to be the case for $z=7-6$. This drop in $f_\mathrm{esc}^\mathrm{ej}$ from $z=7-6$ for $M_h\lesssim10^{9.7}\msun$ can be explained by the non-continuously increasing time steps in this redshift range that our delayed SN feedback scheme is sensitive to (see Appendix \ref{app_subsec_oscillations} for more details).
We would expect that a finer time resolution would result in a smoother evolution of $f_\mathrm{esc}$, with the true values lying within the values demarcated by the oscillations (see Fig. \ref{fig_Nion_binnedMvir_contribution} in Appendix \ref{app_subsec_oscillations}).

We now discuss the environmental dependence of $f_\mathrm{esc}^\mathrm{ej}$ as shown in Fig. \ref{fig_fesc_dens_Mvir_FESCTEMPEVOL}. We find that for low-mass galaxies ($M_h\lesssim10^{10}\msun$), $f_\mathrm{esc}^\mathrm{ej}$ is higher in over-dense than in under-dense regions as it is linked to the SN feedback limited star formation efficiency $f_\star^\mathrm{ej}$.
We can understand the dependence of $f_\star^\mathrm{ej}$ on the underlying density by analysing equation \ref{eq_fej_delayedSN}. While the first factor, $v_c^2/(v_c^2+f_w E_{51} \nu_z)$, depends on the halo mass alone and hence can not cause the observed trend, the second factor, $\sum_{j}\nu_j M_{\star,j}^\mathrm{new}/M_\mathrm{g}^i(z)$, is highly sensitive to the star formation and gas accretion histories of galaxies. The more stars have formed in previous time steps or the less gas is available for star formation in the current time step, the lower is the fraction of gas that needs to form stars to eject all remaining gas from the galaxy ($f_\star^\mathrm{ej}$) and the higher is the fraction of gas that is ejected and correspondingly the ionizing escape fraction $f_\mathrm{esc}^\mathrm{ej}$. Hence, in over-dense regions galaxies of the same halo mass must have lower 
values for the ratio $\sum_{j}\nu_j M_{\star,j}^\mathrm{new}/M_\mathrm{g}^i(z)$, which is due to the following two effects: 
Firstly, the initial gas mass $M_\mathrm{g}^i(z)$ in lower mass galaxies ($M_h\lesssim10^{10}\msun$) is dominated by the accreted gas mass, $M_\mathrm{acc}(z)$, since all or most of the gas is ejected through SN feedback at most time steps. Secondly, since, in hierarchical structure formation, density peaks in over-dense regions collapse prior to those in under-dense regions, galaxies of a given halo mass, $M_h$, and at a given redshift $z$ have formed earlier in over-dense than in under-dense regions and hence show on average lower DM and gas accretion rates due to their longer lifetime. Their lower gas accretion rates result indeed in both, less stellar mass having been formed in previous time steps and exploding as SN at redshift $z$, and less gas being available for star formation at redshift $z$. However, for a fixed time step, the ratio between the two, i.e.  $\sum_{j}\nu_j M_{\star,j}^\mathrm{new}/M_\mathrm{g}^i(z)$, increases towards lower gas accretion rates and leads to a lower value for $f_\star^\mathrm{ej}$.
This trend can be also understood in descriptive terms: The presence of more massive galaxies in over-dense regions leads to unbound gas being more likely to be gravitationally attracted to those more massive galaxies than to low-mass galaxies. We remark that due to the time-delay of the onset of radiative feedback upon reionization, $f_\star^\mathrm{ej}$ is hardly affected by radiative feedback during reionization.

Finally, we see that an ionizing escape fraction that scales with the ejected gas fraction from the galaxy shifts the ionizing emissivity contribution to both lower halo masses (comparing the top left panels of Figs. \ref{fig_Nion_dens_Mvir_SOBACCHI} and \ref{fig_Nion_dens_Mvir_FESCTEMPEVOL}) and UV luminosities (comparing the top left panels of Figs. \ref{fig_Nion_MUV_Ngal_SOBACCHI} and \ref{fig_Nion_MUV_Ngal_FESCTEMPEVOL}). For example, while for the {\it Photoionization} model the ionizing emissivity peaks at $M_h\simeq10^{10}\msun$ at $z=7$, it peaks at a much lower mass of $M_h\simeq10^{9}\msun$ for the {\it Early Heating} model. Similarly, the peak in the ionizing emissivity shifts from $M_\mathrm{UV}\simeq-18$ at $z=7$ in the {\it Photoionization} model to a much lower value of $M_\mathrm{UV}\simeq-14$ in the {\it Early Heating} model.
As a consequence of the ionizing emissivity being shifted to lower mass halos and UV luminosities, the majority of the ionizing emissivity no longer originates from galaxies on the ``halo mass - underlying density" relation described in the previous Section. Instead, the bulk of the ionizing emissivity comes from lower-mass galaxies with $M_h \lesssim 10^{9.5}\msun$) in intermediate- to less-density regions ($\log_{10}(1+\delta)\simeq0-0.5$) (c.f. bottom left panels of Figs. \ref{fig_Nion_MUV_Ngal_SOBACCHI} and \ref{fig_Nion_dens_Mvir_FESCTEMPEVOL})\footnote{In denser regions with $\log_{10}(1+\delta)\simeq0.6-1.2$ ($\log_{10}(N_\mathrm{gal})\simeq(-0.3)-0.2$) the bimodal distribution across halo mass (UV luminosity) with maxima around $M_h\gtrsim10^{9.6}\msun$ ($M_\mathrm{UV}\simeq-15$) and $M_h\simeq10^{11}\msun$ ($M_\mathrm{UV}\lesssim-19$) is due to $f_\mathrm{esc}^\mathrm{ej}$ approaching $f_\mathrm{esc}^0=0.6$ at higher $M_h$ (lower $M_\mathrm{UV}$) values in denser than in less dense regions (see Fig. \ref{fig_fesc_dens_Mvir_FESCTEMPEVOL}). Indeed, the majority of galaxies in denser regions reaches $f_\mathrm{esc}^\mathrm{ej}\simeq f_\mathrm{esc}^0$ at $M_h\simeq10^{9.6}\msun$.}. This redistribution of ionizing photons results in more similar sized ionized (\HII) regions around both massive and low-mass halos as compared to our models using a constant $f_\mathrm{esc}$ value \citep[see also][]{hutter2021}.

We now briefly compare our results to other works. We find our $f_\mathrm{esc}^\mathrm{ej}$-$M_h$ relation to be in good agreement with the trends found in radiation hydrodynamical simulations \citep[e.g.][]{lewis2020, Kimm2014} in that $f_\mathrm{esc}$ decreases with increasing halo mass and increases with decreasing redshift over the entire galaxy population as low mass galaxies become less dense and exhibit lower optical depths for \HI ionizing photons. Furthermore, despite not tracking the inhomogeneous gas distribution within and around galaxies, our analytic model finds similar $f_\mathrm{esc}$ values, particularly towards higher halo masses, $M_h\gtrsim10^{9.5}\msun$.
Finally, we find the evolution of our ionizing emissivity as a function of halo mass to be very similar to the escape emissivity in \citet{lewis2020}, particularly when comparing the shift of the peak and its position with respect to the halo mass.

\section{Quantifying the galaxies contributing to reionization}
\label{sec_ionizing_budget}

\begin{figure}
    \centering
    \includegraphics[width=0.46\textwidth]{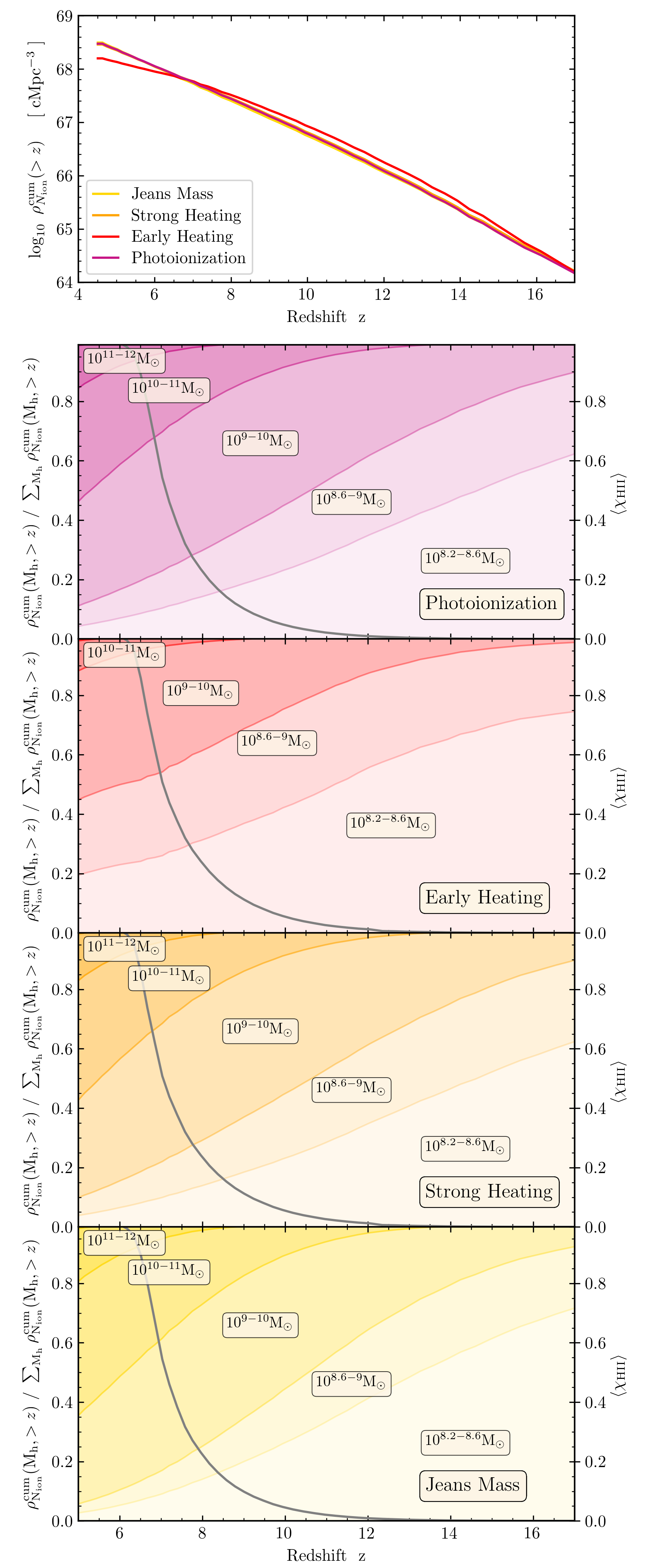}
    \caption{Top: Evolution of the cumulative density of the total number of ionizing photons at redshift $z$ for the {\it Photoionization} (violet), {\it Early Heating} (red), {\it Strong Heating} (orange) and {\it Jeans Mass} model (yellow). Bottom: Fraction of the ionizing photon contribution to reionization from galaxies with halo masses within the indicated range at redshift $z$ for the same models as in the top panel. From light to darker shades the halo mass increases. The gray solid line indicates the evolution of the volume-averaged hydrogen ionization fraction.}
    \label{fig_cumNion_binnedMvir}
\end{figure}

Finally, we quantify the contribution of galaxies of different masses to reionization for all of the radiative feedback models discussed above. We start by briefly commenting on the evolution of the cumulative number density of ionizing photons until redshift $z$, as shown in the top panel of Fig. \ref{fig_cumNion_binnedMvir}. Models with a constant ionizing escape fraction  ({\it Photoionization}, {\it Strong Heating}, {\it Jeans Mass} models) show a very similar evolution, which is also reflected in the strong agreement of their ionization histories (gray solid lines in the bottom panels of Fig. \ref{fig_cumNion_binnedMvir}). However, the evolution of the cumulative number density of ionizing photons for the {\it Early Heating} model (that uses a mass-dependent escape fraction) differs from all the other models, with more (less) ionizing photons being contributed before (after) the midpoint of reionization. These differences indicate two points: firstly, in this scenario, reionization is driven by lower mass galaxies compared to the other models. Secondly, the introduction of a mass-dependent escape fraction seems to have a stronger impact on the galaxies driving reionization as compared to the impact of different radiative feedback prescriptions.

We investigate these findings in detail by analysing the contribution of different halo masses to the ionizing budget from $z\simeq25$ down to redshift $z$, as shown in the bottom panels of Fig. \ref{fig_cumNion_binnedMvir}. Similar to the evolution of the cumulative number density of ionizing photons, all radiative feedback models that assume a constant $f_\mathrm{esc}$ value ({\it Photoionization}, {\it Strong Heating}, and {\it Jeans Mass} models) show very similar contributions of different halo masses to reionization. Before the Universe is $\sim0.5$\% ionized at $z\simeq13$, low-mass galaxies ($M_h<10^9\msun$) located in slightly over-dense regions ($\log_{10}(1+\delta)\simeq0-0.4$) contribute more than 60\% to the total ionizing budget. Slightly more massive galaxies ($M_h\gtrsim10^{9-10}\msun$) located in intermediate dense regions ($\log_{10}(1+\delta)\simeq0.1-0.4$) are less affected by SN feedback, resulting in smoother star formation histories, and are able to build up ionized regions of about $\sim1-2$~Mpc. The continually growing mass (and ionizing output) of these galaxies increases with decreasing redshift. By the mid-point of reionization, at $z\simeq7.2$, galaxies with $M_h\sim 10^{9}\msun$ located in intermediate dense to dense regions ($\log_{10}(1+\delta)\simeq0.1-0.8$) and corresponding to those with mostly continuous star formation histories, have provided more than $75$\% of the ionizing photons even for our minimum radiative feedback scenario (see {\it Minimum} model) in \citet{hutter2021}. Increasing the radiative feedback strength, i.e. going to the {\it Strong Heating} and {\it Jeans Mass} models, results in an even higher contribution of $>80$\%, given that radiative feedback suppresses the contribution of lower-mass halos ($M_h<10^{10}\msun$) at all redshifts. For all constant $f_\mathrm{esc}$ models we find the ionizing photon contribution to be dominated by $M_h=10^{9-10}\msun$ halos during the first half of reionization and by $M_h=10^{9-10}\msun$ and increasingly $M_h=10^{10-11}\msun$ halos during the second half of reionization.
However, introducing a biased ionizing escape fraction changes this picture considerably. In our {\it Early Heating} model, we find low-mass galaxies located in slightly dense to intermediate dense regions ($\log_{10}(1+\delta)\simeq0-0.5$) to be the largest contributors to reionization: at $z\simeq13$ they have provided $\simeq90$\% of the ionizing photons, and even at the midpoint of reionization their cumulative contribution adds up to $\simeq50-55$\%.

Finally, we compare our results for our {\it Early Heating} model to those obtained with radiation hydrodynamical simulations \citep{lewis2020, katz2018, Kimm2014} and hydrodynamical simulations post-processed with radiative transfer \citep{yajima2011}.
In case of \citet{lewis2020}, we note two differences that cause a change in the galaxy population that contributes the ionizing photons. Firstly, their $f_\mathrm{esc}-M_h$ relation is less steep, i.e. low-mass halos have lower and high-mass halos have higher $f_\mathrm{esc}$ values. Secondly, in {\sc CoDa ii} {\it only} the youngest stellar populations ($<10$~Myr) contribute to the ionizing emissivity, while we also consider the contribution from older stellar populations; hence their peak of the total ionizing emissivity is shifted to higher halo masses and the contribution of galaxies in low-mass halos is lower than in our simulations. Both effects result in low-mass halos ($M_h\lesssim10^9\msun$) contributing far less to reionization as compared to our corresponding {\it Early Heating} simulation ($\sim20$\% at $z\simeq7$ in \citet{lewis2020} compared to $\sim40$\% in our simulations as seen in Fig. \ref{fig_Nion_binnedMvir_contribution}), while the contribution from more massive galaxies with $M_h\gtrsim10^9\msun$ is higher, being $\sim60$\% ($\sim20$\%) in \citet{lewis2020} for $M_h=10^{9-10}\msun$ ($M_h=10^{10-11}\msun$) compared to $50$\% ($10$\%) in our simulation as seen in Fig. \ref{fig_Nion_binnedMvir_contribution}). \citet{katz2018} find an even stronger contribution from higher mass galaxies than \citet{lewis2020} and our {\it Early Heating} simulation ($\sim10$\% for $M_h<10^{9.2}\msun$, $\sim40$\% for $10^{9.2-10.2}\msun$ and $\sim50$\% for $M_h>10^{10.2}\msun$ at $z=7$), which is partly driven by the flatter $f_\mathrm{esc}-M_h$ relation with low-mass (higher mass) halos having even lower (higher) escape fractions as compared to our model. For a similar reason, the contribution of massive galaxies ($M_h>10^{10}\msun$) to the cumulative number of ionizing photons up to $z=7$ in \citet{Kimm2014} lies between the results of our {\it Early Heating} simulation and simulations assuming a constant $f_\mathrm{esc}$ value.
On the other hand, \citet{yajima2011} obtain results in line with ours: i.e. a strong contribution from low-mass galaxies of about $75$\% to the ionizing photon density compared to $\sim80$\% in our models (c.f. Fig. \ref{fig_Nion_binnedMvir_contribution} in Appendix \ref{app_ionizing_emissivity}) at $z=6$. 

\begin{table*}
    \centering
    \begin{tabular}{c|c|c|c|c|c}
    \hline
         $f_\mathrm{nion}(z=10)$ & $M_\mathrm{UV}\leq-13$ & $M_\mathrm{UV}\leq-14$ & $M_\mathrm{UV}\leq-15$ &$M_\mathrm{UV}\leq-16$ & $M_\mathrm{UV}\leq-17$\\
         \hline
         \hline
         {\it Photoionization} & 0.83 & 0.72 & 0.58 & 0.43 & 0.27 \\
         \hline
         {\it Early Heating} & 0.63 & 0.42 & 0.21 & 0.08 & 0.03 \\
         \hline
         {\it Strong Heating} & 0.83 & 0.72 & 0.59 & 0.44 & 0.28 \\
         \hline
         {\it Jeans Mass} & 0.75 & 0.67 & 0.59 & 0.47 & 0.3\\
         \hline
         \\
         \hline
         $f_\mathrm{nion}(z=7)$ & $M_\mathrm{UV}\leq-13$ & $M_\mathrm{UV}\leq-14$ & $M_\mathrm{UV}\leq-15$ &$M_\mathrm{UV}\leq-16$ & $M_\mathrm{UV}\leq-17$\\
         \hline
         \hline
         {\it Photoionization} & 0.86 & 0.78 & 0.69 & 0.58 & 0.46 \\
         \hline
         {\it Early Heating} & 0.57 & 0.37 & 0.21 & 0.12 & 0.07 \\
         \hline
         {\it Strong Heating} & 0.86 & 0.79 & 0.71 & 0.60 & 0.48 \\
         \hline
         {\it Jeans Mass} & 0.90 & 0.86 & 0.80 & 0.70 & 0.56 \\
         \hline
    \end{tabular}
    \caption{The fraction $f_\mathrm{nion}$ of the total ionizing emissivity at $z=10$ (top) and $z=7$ (bottom) that comes from galaxies with a minimum UV luminosity, $M_\mathrm{UV}$. We show the fractions for the {\it Photoionization}, {\it Early Heating}, {\it Strong Heating} and {\it Jeans Mass} models.}
    \label{tab_fnion_MUV}
\end{table*}

\section{Conclusions}
\label{sec_conclusions}

\begin{figure}
    \centering
    \includegraphics[width=0.49\textwidth]{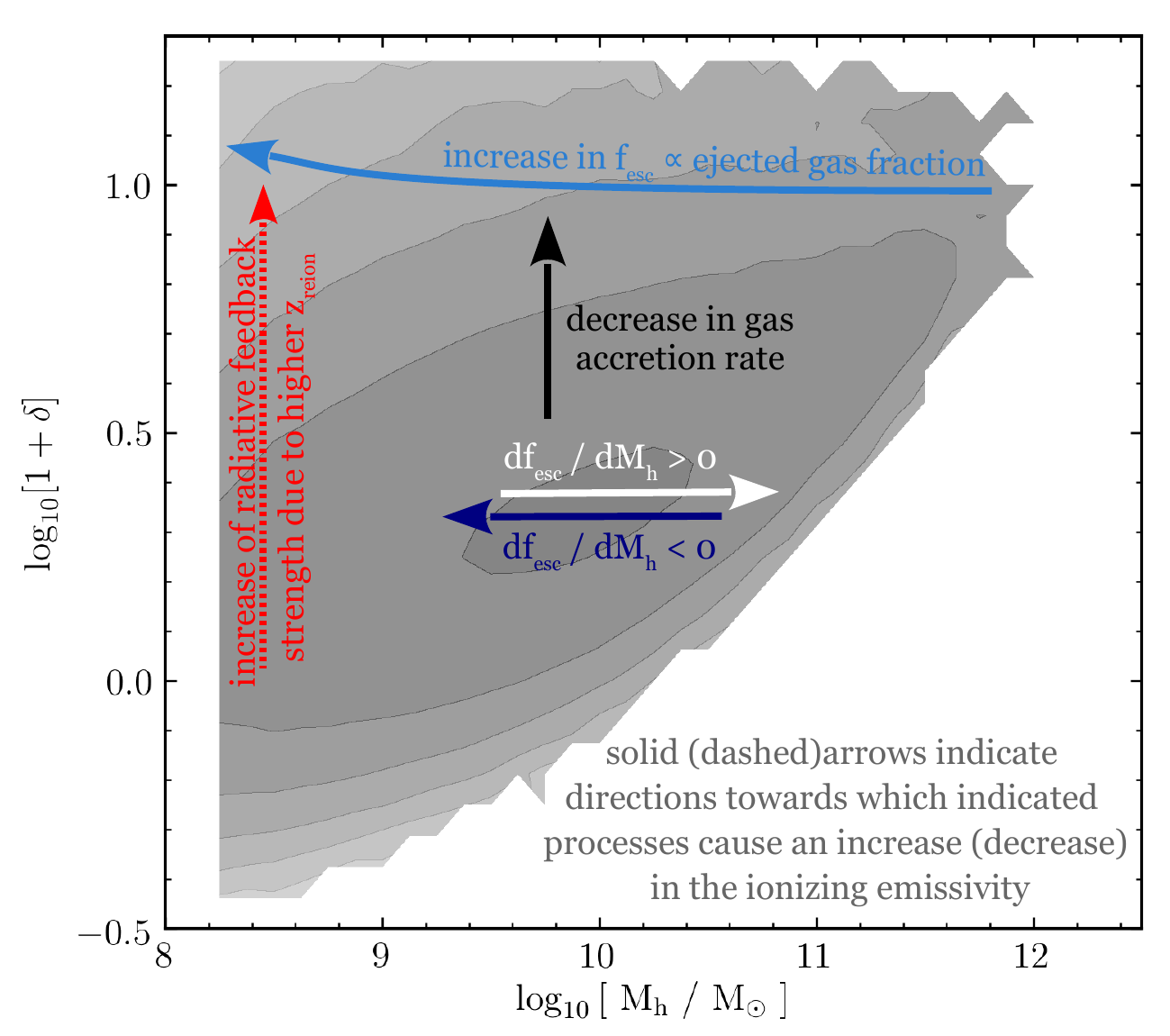}
    \caption{Schematic illustrating how different processes affect the ionizing emissivity contribution from different halo masses. Grey contours indicate the ionizing emissivity distribution as a function of halo mass and the over-density smoothed over $2$~cMpc for a model without radiative feedback. Solid (dashed) arrows indicate the direction in which the marked processes lead to an increase (decrease) in the ionizing emissivity contribution.}
    \label{fig_Nion_dens_Mvir_cartoon}
\end{figure}

In this work, our aim was to investigate the impact of physical properties (such as the escape fraction of ionizing photons) as well as the environment (density and ionization fields) in determining the key reionization sources. For this purpose we have used the {\sc astraeus} framework that self-consistently couples galaxy formation and reionization using an N-body simulation ({\sc very small multidark planck; vsmdpl}), a semi-analytic model for galaxy formation (an extended version of {\sc Delphi} and a seminumerical radiatve tranfer code for reionization {\sc cifog}). The key strength of our model lies in the range of escape fraction ($f_\mathrm{esc}$) and radiative feedback scenarios explored. We include two different models for the escape fraction, one where $f_\mathrm{esc}$ remains constant for all galaxies at all redshifts ({\it Photoionization}, {\it Strong Heating} and {\it Jeans Mass} model), and one where $f_\mathrm{esc}$ scales with the gas fraction that is ejected from the galaxy (increasing with decreasing halo mass, {\it Early Heating} model). Additionally, our radiative feedback scenarios range from a weak and time-delayed model ({\it Photoionization} model) to a strong and immediate reduction in the gas content and star formation of a galaxy ({\it Jeans Mass} model). Our main findings are:

\begin{enumerate}
    \item When assuming a constant ionizing escape fraction $f_\mathrm{esc}$, intermediate mass ($M_h\simeq10^{9-11}$) and relatively luminous galaxies ($M_\mathrm{UV}\simeq-15$ to $-20$) in average- to intermediate-density regions ($\log_{10}(1+\delta) \simeq 0-0.8$) are the key drivers of reionization. 
    \item While the mass assembly histories of low- and intermediate-mass galaxies ($M_h\lesssim10^{8.5-11}\msun$) in over-dense ($\log_{10}(1+\delta)\gtrsim 0.2$ for $M_h\simeq10^9\msun$, and $\log_{10}(1+\delta)\gtrsim 1$ for $M_h\simeq10^{11}\msun$) and under-dense regions ($\log_{10}(1+\delta)\lesssim0.2$ for $M_h\simeq10^9\msun$, and $\log_{10}(1+\delta)\lesssim 1$ for $M_h\simeq10^{11}\msun$) differ only marginally and hardly affect their ionizing emissivities in different environments (see black arrow in Fig. \ref{fig_Nion_dens_Mvir_cartoon} for trend), reionization imprints its topology on galaxy formation through radiative feedback effects on low-mass galaxies ($M_h\lesssim10^{9}\msun$). With over-dense regions being reionized before under-dense regions, star formation in low-mass galaxies is more suppressed (by time-delayed radiative feedback) in over-dense regions, resulting in a lower ionizing emissivity (see red arrow in Fig. \ref{fig_Nion_dens_Mvir_cartoon}).
    \item Linking the ionizing escape fraction  to the gas fraction ejected from the galaxy results in an $f_\mathrm{esc}$ that decreases with increasing halo mass. Additionally, such an $f_\mathrm{esc}$ shows an increase with decreasing redshift for a given halo mass (see blue and green arrows in Fig. \ref{fig_Nion_dens_Mvir_cartoon}). Furthermore, this escape fraction shows a slight dependence on the environment due to small differences in the star formation and gas accretion histories of low and intermediate mass galaxies in over- versus under-dense regions. With galaxies in over-dense regions forming earlier and showing lower gas accretion rates, $f_\mathrm{esc}$ has higher values in over-dense than in under-dense regions (see top light blue arrow in Fig. \ref{fig_Nion_dens_Mvir_cartoon}). 
    \item Assuming an ionizing escape fraction that decreases with increasing halo mass results in low-mass galaxies ($M_h\lesssim10^{9.5}\msun$) in less-dense regions ($\log_{10}(1+\delta)\simeq0-0.5$) being the main drivers of reionization.
    \item As seen from Fig. \ref{fig_Nion_dens_Mvir_cartoon}, increasing radiative feedback strength reduces the contribution of low-mass galaxies in dense regions to reionization. For our {\it Photoionization} model, radiative feedback decreases the ionizing emissivity of the lowest-mass galaxies ($M_h\lesssim10^{8.5}\msun$) by an order of magnitude when going from average ($\log_{10}(1+\delta)\simeq0$) to over-dense regions ($\log_{10}(1+\delta)\simeq1$).
    \end{enumerate}

There are a few caveats to our work.
Firstly, our underlying N-body simulation resolves halos down to $M_h\simeq10^{8.2}\msun$, hence we can not quantify the contribution of even lower-mass halos and the initial star formation histories of halos with $M_h<10^{8.6}\msun$ might have not converged yet \citep[see Appendix B in][]{hutter2021}. However, when assuming a constant ionizing escape fraction, the limited resolution should hardly affect our results, since lower mass halos will be even more subject to SN and radiative feedback and their contribution will be similarly or even more limited as that of $M_h=10^{8.2-9}\msun$ halos\footnote{A rough estimate, assuming a constant gas-to-DM mass fraction and instantaneous SN feedback ($f_\star^\mathrm{ej}=v_c^2 / (v_c^2 E_{51} \nu_z)$), yields that the newly stellar mass formed in $10^7-10^8\msun$ halos is at most $25-35\%$ of that formed in $10^8-10^9\msun$ halos at $z\simeq6-9$. In reality, the gas-to-DM mass fraction will be lower for lower mass halos as a result of SN feedback \citep[see][]{hutter2021}, making their contribution to newly formed stellar mass and hence the ionizing emissivity even lower.} (see Appendix \ref{app_reion_contribution_minihalos} for an ionizing budget analysis when running our model on a N-body simulation with a $20\times$ higher mass resolution).
Nevertheless, such lower mass halos might be crucial for scenarios where the escape fraction decreases with halo mass. Their contribution might be enhanced due to the higher $f_\mathrm{esc}$ values for lower halo masses but could also be insignificant due to being strongly suppressed by SN and radiative feedback. 
Indeed, recent radiation hydrodynamical simulations hint at that - even for ionizing escape fractions increasing with decreasing halo mass - $10^7-10^8\msun$ halos hardly contribute to reionizing the Universe at $z\lesssim9-10$ \citep{Kimm2017, norman2018}. \citet{norman2018} find that $10^7-10^8\msun$ halos contribute more than $10^8-10^9\msun$, however they do not account for the increasing effects of radiative feedback from more massive galaxies and their choice of ionizing emissivities for $10^8-10^9\msun$ halos underestimates the contribution of those at $z\gtrsim10$.
Hence, despite not following the contribution from $\lsim10^8\msun$ halos, our simulations strongly indicate that the escape fraction plays a crucial role in which galaxy population is the main contributor to reionization.

Secondly, since our simulations do not follow the formation and destruction of dust, the UV luminosities of our simulated galaxies are not dust attenuated. However, dust is only expected to impact the UV luminosities for galaxies with $M_\mathrm{UV}\lesssim-18$, showing an average shift of $\Delta M_\mathrm{UV}\simeq0.5$ to $1$ at $M_\mathrm{UV}\simeq-20$ to $-21$ \citep{garel2015, yung2019, yung2020}. Given that such galaxies contribute only $\sim20\%$ to reionization by $z\sim6$, the inclusion of dust is not expected to affect our results in any sensible way. We caution that the impact of dust on the measured rest-frame UV continuum spectral index is not easy to interpret, being degenerate with the age of the stellar population, its initial mass function (IMF) and binary fractions. Additionally, due to a lack of very deep infrared observations, the dust masses and temperatures of early galaxies remain highly debated \citep[e.g.][]{mancini2016, behrens2018}.

Thirdly, we assume a universal IMF, which might be particularly important at the beginning of reionization and for the ionizing emissivity of low-mass halos. Due to their low star formation rates low-mass halos might have a more bottom-heavy IMF \citep{jerabkova2018}, which would cause them to transition earlier, i.e. at lower halo masses, from stochastic to continuous star formation and hence to enhance their contribution to reionization. Alternatively, if due to their lower metallicity and starburst character low-mass halos had a more top-heavy IMF, they would contribute even less to the ionizing budget for reionization. However, galaxies in low-mass halos in over-dense regions may have a higher metallicity, which would boost again their contribution to reionization compared to the metal-poor low-mass galaxies in under-dense regions. How strong the effect of a spatially varying metallicity on the production of ionizing photons and whether an ionizing escape fraction that scales with the ejected gas fraction would compensate this effect remains open and will be subject of future work.

Imaging surveys with the JWST will greatly enhance the number of observed high-redshift galaxies down to $M_\mathrm{UV}\simeq-16$ at $z\simeq7$ \citep[see][for the assumed limit]{williams2018}. In our models, the contribution of galaxies detected at $z\simeq7$ to the total ionizing photon budget at $z\simeq7$ varies between $\sim10\%$ and $\sim70\%$, with the exact value being strongly dependent on the mass dependence of $f_\mathrm{esc}$ and the radiative feedback strength suppressing star formation in low-mass halos and determining the turn-over of the UV luminosity function at its faint end (c.f. Table \ref{tab_fnion_MUV}).
In addition, the Square Kilometre Array (SKA) will detect the 21cm signal from the neutral hydrogen in the IGM mapping the time and spatial evolution of the ionized regions around galaxies. Combining JWST's and ALMA's unprecedented UV luminosity and emission line observations from galaxies with the SKA's measurements of the ionization topology will help us to pin down the properties of the star-forming galaxies, such as their stellar populations and ionizing escape fractions, and constrain the main contributors to reionization. 

\section*{Acknowledgements} 

The authors thank the anonymous referee for their comments.
All the authors acknowledge support from the European Research Council's starting grant ERC StG-717001 (``DELPHI"). PD acknowledges support from the NWO grant 016.VIDI.189.162 (``ODIN") and the European Commission's and University of Groningen's CO-FUND Rosalind Franklin program. 
GY acknowledges financial support from  MICIU/FEDER under project grant PGC2018-094975-C21.
The authors wish to thank Peter Behroozi for creating and providing the {\sc rockstar} merger trees of the VSMPL and ESMDPL simulations, and V. Springel for allowing us to use the L-Gadget2 code to run the different Multidark simulation boxes, including the VSMDPL and ESMDPL used in this work. The VSMDPL and ESMDPL simulations have been performed at LRZ Munich within the project pr87yi. The CosmoSim database (\url{www.cosmosim.org}) provides access to the simulation and the Rockstar data. The database is a service by the Leibniz Institute for Astrophysics Potsdam (AIP). 

This research made use of \texttt{matplotlib}, a Python library for publication quality graphics \citep{hunter2007}; and the Python library \texttt{numpy} \citep{numpy}.

\section*{Data Availability} 

The source code of the seminumerical galaxy evolution and reionization model within the {\sc astraeus} framework and the employed analysis scripts are available on GitHub (\url{https://github.com/annehutter/astraeus}). The underlying N-body DM simulation, the {\sc astraeus} simulations and derived data in this research will be shared on reasonable request to the corresponding author.

\bibliographystyle{mn2e}
\bibliography{delphi}

\begin{thebibliography}{57}
\expandafter\ifx\csname natexlab\endcsname\relax\def\natexlab#1{#1}\fi

\bibitem[{{Abel} {et~al}\mbox{.}(2007){Abel}, {Wise}, \& {Bryan}}]{abel2007}
{Abel} T., {Wise} J.~H., {Bryan} G.~L., 2007, \apjl, 659, L87

\bibitem[{{Barkana} \& {Loeb}(1999)}]{barkana-loeb1999}
{Barkana} R., {Loeb} A., 1999, \apj, 523, 54

\bibitem[{{Behrens} {et~al}\mbox{.}(2018){Behrens}, {Pallottini}, {Ferrara},
  {Gallerani}, \& {Vallini}}]{behrens2018}
{Behrens} C., {Pallottini} A., {Ferrara} A., {Gallerani} S., {Vallini} L.,
  2018, \mnras, 477, 552

\bibitem[{{Behroozi} {et~al}\mbox{.}(2013{\natexlab{a}}){Behroozi}, {Wechsler},
  \& {Wu}}]{behroozi2013_rs}
{Behroozi} P.~S., {Wechsler} R.~H., {Wu} H.-Y., 2013{\natexlab{a}}, \apj, 762,
  109

\bibitem[{{Behroozi} {et~al}\mbox{.}(2013{\natexlab{b}}){Behroozi}, {Wechsler},
  {Wu}, {Busha}, {Klypin}, \& {Primack}}]{behroozi2013_trees}
{Behroozi} P.~S., {Wechsler} R.~H., {Wu} H.-Y., {Busha} M.~T., {Klypin} A.~A.,
  {Primack} J.~R., 2013{\natexlab{b}}, \apj, 763, 18

\bibitem[{{Bhatawdekar} {et~al}\mbox{.}(2019){Bhatawdekar}, {Conselice},
  {Margalef-Bentabol}, \& {Duncan}}]{bhatawdekar2019}
{Bhatawdekar} R., {Conselice} C.~J., {Margalef-Bentabol} B., {Duncan} K., 2019,
  \mnras, 486, 3805

\bibitem[{{Bouwens} {et~al}\mbox{.}(2015){Bouwens}, {Smit}, {Labbe}, {Franx},
  {Caruana}, {Oesch}, {Stefanon}, \& {Rasappu}}]{bouwens2015b}
{Bouwens} R.~J., {Smit} R., {Labbe} I., {Franx} M., {Caruana} J., {Oesch} P.,
  {Stefanon} M., {Rasappu} N., 2015, ArXiv e-prints

\bibitem[{{Bouwens} {et~al}\mbox{.}(2019){Bouwens}, {Stefanon}, {Oesch},
  {Illingworth}, {Nanayakkara}, {Roberts-Borsani}, {Labb{\'e}}, \&
  {Smit}}]{bouwens2019}
{Bouwens} R.~J., {Stefanon} M., {Oesch} P.~A., {Illingworth} G.~D.,
  {Nanayakkara} T., {Roberts-Borsani} G., {Labb{\'e}} I., {Smit} R., 2019,
  \apj, 880, 25

\bibitem[{{Couchman} \& {Rees}(1986)}]{couchman-rees1986}
{Couchman} H.~M.~P., {Rees} M.~J., 1986, \mnras, 221, 53

\bibitem[{{Davies} {et~al}\mbox{.}(2018){Davies}, {Hennawi}, {Ba{\~n}ados},
  {Luki{\'c}}, {Decarli}, {Fan}, {Farina}, {Mazzucchelli}, {Rix}, {Venemans},
  {Walter}, {Wang}, \& {Yang}}]{davies2018}
{Davies} F.~B. {et~al.}, 2018, \apj, 864, 142

\bibitem[{{Dayal} \& {Ferrara}(2018)}]{dayal2018}
{Dayal} P., {Ferrara} A., 2018, \physrep, 780, 1

\bibitem[{{Dayal} {et~al}\mbox{.}(2014){Dayal}, {Ferrara}, {Dunlop}, \&
  {Pacucci}}]{dayal2014}
{Dayal} P., {Ferrara} A., {Dunlop} J.~S., {Pacucci} F., 2014, \mnras, 445, 2545

\bibitem[{{Dayal} {et~al}\mbox{.}(2020){Dayal}, {Volonteri}, {Choudhury},
  {Schneider}, {Trebitsch}, {Gnedin}, {Atek}, {Hirschmann}, \&
  {Reines}}]{dayal2020}
{Dayal} P. {et~al.}, 2020, \mnras, 495, 3065

\bibitem[{{Fan} {et~al}\mbox{.}(2006){Fan}, {Strauss}, {Becker}, {White},
  {Gunn}, {Knapp}, {Richards}, {Schneider}, {Brinkmann}, \&
  {Fukugita}}]{fan2006}
{Fan} X. {et~al.}, 2006, \aj, 132, 117

\bibitem[{{Furlanetto} {et~al}\mbox{.}(2004){Furlanetto}, {Zaldarriaga}, \&
  {Hernquist}}]{furlanetto2004}
{Furlanetto} S.~R., {Zaldarriaga} M., {Hernquist} L., 2004, \apj, 613, 1

\bibitem[{{Garel} {et~al}\mbox{.}(2015){Garel}, {Blaizot}, {Guiderdoni},
  {Michel-Dansac}, {Hayes}, \& {Verhamme}}]{garel2015}
{Garel} T., {Blaizot} J., {Guiderdoni} B., {Michel-Dansac} L., {Hayes} M.,
  {Verhamme} A., 2015, \mnras, 450, 1279

\bibitem[{{Gnedin}(2000)}]{gnedin2000}
{Gnedin} N.~Y., 2000, \apj, 542, 535

\bibitem[{{Gnedin} \& {Hui}(1998)}]{gnedin1998b}
{Gnedin} N.~Y., {Hui} L., 1998, \mnras, 296, 44

\bibitem[{{Hassan} {et~al}\mbox{.}(2018){Hassan}, {Dav{\'e}}, {Mitra},
  {Finlator}, {Ciardi}, \& {Santos}}]{hassan2018}
{Hassan} S., {Dav{\'e}} R., {Mitra} S., {Finlator} K., {Ciardi} B., {Santos}
  M.~G., 2018, \mnras, 473, 227

\bibitem[{{Hoeft} {et~al}\mbox{.}(2006){Hoeft}, {Yepes}, {Gottl{\"o}ber}, \&
  {Springel}}]{hoeft2006}
{Hoeft} M., {Yepes} G., {Gottl{\"o}ber} S., {Springel} V., 2006, \mnras, 371,
  401

\bibitem[{Hunter(2007)}]{hunter2007}
Hunter J.~D., 2007, Computing In Science \& Engineering, 9, 90

\bibitem[{{Hutter}(2018)}]{hutter2018}
{Hutter} A., 2018, \mnras, 477, 1549

\bibitem[{{Hutter} {et~al}\mbox{.}(2021){Hutter}, {Dayal}, {Yepes},
  {Gottl{\"o}ber}, {Legrand}, \& {Ucci}}]{hutter2021}
{Hutter} A., {Dayal} P., {Yepes} G., {Gottl{\"o}ber} S., {Legrand} L., {Ucci}
  G., 2021, \mnras

\bibitem[{{Ishigaki} {et~al}\mbox{.}(2018){Ishigaki}, {Kawamata}, {Ouchi},
  {Oguri}, {Shimasaku}, \& {Ono}}]{ishigaki2018}
{Ishigaki} M., {Kawamata} R., {Ouchi} M., {Oguri} M., {Shimasaku} K., {Ono} Y.,
  2018, \apj, 854, 73

\bibitem[{{Je{\v{r}}{\'a}bkov{\'a}}
  {et~al}\mbox{.}(2018){Je{\v{r}}{\'a}bkov{\'a}}, {Hasani Zonoozi}, {Kroupa},
  {Beccari}, {Yan}, {Vazdekis}, \& {Zhang}}]{jerabkova2018}
{Je{\v{r}}{\'a}bkov{\'a}} T., {Hasani Zonoozi} A., {Kroupa} P., {Beccari} G.,
  {Yan} Z., {Vazdekis} A., {Zhang} Z.~Y., 2018, \aap, 620, A39

\bibitem[{{Katz} {et~al}\mbox{.}(2018){Katz}, {Kimm}, {Haehnelt}, {Sijacki},
  {Rosdahl}, \& {Blaizot}}]{katz2018}
{Katz} H., {Kimm} T., {Haehnelt} M., {Sijacki} D., {Rosdahl} J., {Blaizot} J.,
  2018, \mnras, 478, 4986

\bibitem[{{Kimm} {et~al}\mbox{.}(2019){Kimm}, {Blaizot}, {Garel},
  {Michel-Dansac}, {Katz}, {Rosdahl}, {Verhamme}, \& {Haehnelt}}]{Kimm2019}
{Kimm} T., {Blaizot} J., {Garel} T., {Michel-Dansac} L., {Katz} H., {Rosdahl}
  J., {Verhamme} A., {Haehnelt} M., 2019, \mnras, 486, 2215

\bibitem[{{Kimm} \& {Cen}(2014)}]{Kimm2014}
{Kimm} T., {Cen} R., 2014, \apj, 788, 121

\bibitem[{{Kimm} {et~al}\mbox{.}(2017){Kimm}, {Katz}, {Haehnelt}, {Rosdahl},
  {Devriendt}, \& {Slyz}}]{Kimm2017}
{Kimm} T., {Katz} H., {Haehnelt} M., {Rosdahl} J., {Devriendt} J., {Slyz} A.,
  2017, \mnras, 466, 4826

\bibitem[{{Kitayama} {et~al}\mbox{.}(2004){Kitayama}, {Yoshida}, {Susa}, \&
  {Umemura}}]{kitayama2004}
{Kitayama} T., {Yoshida} N., {Susa} H., {Umemura} M., 2004, \apj, 613, 631

\bibitem[{{Klypin} {et~al}\mbox{.}(2016){Klypin}, {Yepes}, {Gottl{\"o}ber},
  {Prada}, \& {He{\ss}}}]{klypin2016}
{Klypin} A., {Yepes} G., {Gottl{\"o}ber} S., {Prada} F., {He{\ss}} S., 2016,
  \mnras, 457, 4340

\bibitem[{{Leitherer} {et~al}\mbox{.}(1999){Leitherer}, {Schaerer}, {Goldader},
  {Gonz{\'a}lez Delgado}, {Robert}, {Kune}, {de Mello}, {Devost}, \&
  {Heckman}}]{leitherer1999}
{Leitherer} C. {et~al.}, 1999, \apjs, 123, 3

\bibitem[{{Lewis} {et~al}\mbox{.}(2020){Lewis}, {Ocvirk}, {Aubert}, {Sorce},
  {Shapiro}, {Deparis}, {Dawoodbhoy}, {Teyssier}, {Yepes}, {Gottl{\"o}ber},
  {Ahn}, {Iliev}, \& {Chardin}}]{lewis2020}
{Lewis} J. S.~W. {et~al.}, 2020, \mnras

\bibitem[{{Mancini} {et~al}\mbox{.}(2016){Mancini}, {Schneider}, {Graziani},
  {Valiante}, {Dayal}, {Maio}, \& {Ciardi}}]{mancini2016}
{Mancini} M., {Schneider} R., {Graziani} L., {Valiante} R., {Dayal} P., {Maio}
  U., {Ciardi} B., 2016, \mnras, 462, 3130

\bibitem[{{Naidu} {et~al}\mbox{.}(2020){Naidu}, {Tacchella}, {Mason}, {Bose},
  {Oesch}, \& {Conroy}}]{naidu2020}
{Naidu} R.~P., {Tacchella} S., {Mason} C.~A., {Bose} S., {Oesch} P.~A.,
  {Conroy} C., 2020, \apj, 892, 109

\bibitem[{{Norman} {et~al}\mbox{.}(2018){Norman}, {Chen}, {Wise}, \&
  {Xu}}]{norman2018}
{Norman} M.~L., {Chen} P., {Wise} J.~H., {Xu} H., 2018, \apj, 867, 27

\bibitem[{Oliphant(2006)}]{numpy}
Oliphant T., 2006, {NumPy}: A guide to {NumPy}. USA: Trelgol Publishing,
  [Online; accessed <today>]

\bibitem[{{Paardekooper} {et~al}\mbox{.}(2015){Paardekooper}, {Khochfar}, \&
  {Dalla Vecchia}}]{paardekooper2015}
{Paardekooper} J.-P., {Khochfar} S., {Dalla Vecchia} C., 2015, \mnras, 451,
  2544

\bibitem[{{Padovani} \& {Matteucci}(1993)}]{padovani1993}
{Padovani} P., {Matteucci} F., 1993, \apj, 416, 26

\bibitem[{{Parsa} {et~al}\mbox{.}(2018){Parsa}, {Dunlop}, \&
  {McLure}}]{parsa2018}
{Parsa} S., {Dunlop} J.~S., {McLure} R.~J., 2018, \mnras, 474, 2904

\bibitem[{{Planck Collaboration} {et~al}\mbox{.}(2018){Planck Collaboration},
  {Aghanim}, {Akrami}, {Ashdown}, {Aumont}, {Baccigalupi}, {Ballardini},
  {Banday}, {Barreiro}, {Bartolo}, {Basak}, {Battye}, {Benabed}, {Bernard},
  {Bersanelli}, {Bielewicz}, {Bock}, {Bond}, {Borrill}, {Bouchet}, {Boulanger},
  {Bucher}, {Burigana}, {Butler}, {Calabrese}, {Cardoso}, {Carron},
  {Challinor}, {Chiang}, {Chluba}, {Colombo}, {Combet}, {Contreras}, {Crill},
  {Cuttaia}, {de Bernardis}, {de Zotti}, {Delabrouille}, {Delouis}, {Di
  Valentino}, {Diego}, {Dor{\'e}}, {Douspis}, {Ducout}, {Dupac}, {Dusini},
  {Efstathiou}, {Elsner}, {En{\ss}lin}, {Eriksen}, {Fantaye}, {Farhang},
  {Fergusson}, {Fernandez-Cobos}, {Finelli}, {Forastieri}, {Frailis},
  {Fraisse}, {Franceschi}, {Frolov}, {Galeotta}, {Galli}, {Ganga},
  {G{\'e}nova-Santos}, {Gerbino}, {Ghosh}, {Gonz{\'a}lez-Nuevo}, {G{\'o}rski},
  {Gratton}, {Gruppuso}, {Gudmundsson}, {Hamann}, {Handley}, {Hansen},
  {Herranz}, {Hildebrandt}, {Hivon}, {Huang}, {Jaffe}, {Jones}, {Karakci},
  {Keih{\"a}nen}, {Keskitalo}, {Kiiveri}, {Kim}, {Kisner}, {Knox},
  {Krachmalnicoff}, {Kunz}, {Kurki-Suonio}, {Lagache}, {Lamarre}, {Lasenby},
  {Lattanzi}, {Lawrence}, {Le Jeune}, {Lemos}, {Lesgourgues}, {Levrier},
  {Lewis}, {Liguori}, {Lilje}, {Lilley}, {Lindholm}, {L{\'o}pez-Caniego},
  {Lubin}, {Ma}, {Mac{\'\i}as-P{\'e}rez}, {Maggio}, {Maino}, {Mandolesi},
  {Mangilli}, {Marcos-Caballero}, {Maris}, {Martin}, {Martinelli},
  {Mart{\'\i}nez-Gonz{\'a}lez}, {Matarrese}, {Mauri}, {McEwen}, {Meinhold},
  {Melchiorri}, {Mennella}, {Migliaccio}, {Millea}, {Mitra},
  {Miville-Desch{\^e}nes}, {Molinari}, {Montier}, {Morgante}, {Moss}, {Natoli},
  {N{\o}rgaard-Nielsen}, {Pagano}, {Paoletti}, {Partridge}, {Patanchon},
  {Peiris}, {Perrotta}, {Pettorino}, {Piacentini}, {Polastri}, {Polenta},
  {Puget}, {Rachen}, {Reinecke}, {Remazeilles}, {Renzi}, {Rocha}, {Rosset},
  {Roudier}, {Rubi{\~n}o-Mart{\'\i}n}, {Ruiz-Granados}, {Salvati}, {Sandri},
  {Savelainen}, {Scott}, {Shellard}, {Sirignano}, {Sirri}, {Spencer},
  {Sunyaev}, {Suur-Uski}, {Tauber}, {Tavagnacco}, {Tenti}, {Toffolatti},
  {Tomasi}, {Trombetti}, {Valenziano}, {Valiviita}, {Van Tent}, {Vibert},
  {Vielva}, {Villa}, {Vittorio}, {Wand elt}, {Wehus}, {White}, {White},
  {Zacchei}, \& {Zonca}}]{planck2018}
{Planck Collaboration} {et~al.}, 2018, arXiv e-prints, arXiv:1807.06209

\bibitem[{{Rieke} {et~al}\mbox{.}(2019){Rieke}, {Arribas}, {Bunker}, {Charlot},
  {Finkelstein}, {Maiolino}, {Robertson}, {Willott}, {Windhorst}, {Eisenstein},
  {Nelson}, {Tacchella}, {Egami}, {Endsley}, {Frye}, {Hainline}, {Hviding},
  {Rieke}, {Williams}, {Willmer}, \& {Woodrum}}]{rieke2019}
{Rieke} M. {et~al.}, 2019, \baas, 51, 45

\bibitem[{{Salpeter}(1955)}]{salpeter1955}
{Salpeter} E.~E., 1955, \apj, 121, 161

\bibitem[{{Seiler} {et~al}\mbox{.}(2018){Seiler}, {Hutter}, {Sinha}, \&
  {Croton}}]{Seiler2018}
{Seiler} J., {Hutter} A., {Sinha} M., {Croton} D., 2018, \mnras, 480, L33

\bibitem[{{Seiler} {et~al}\mbox{.}(2019){Seiler}, {Hutter}, {Sinha}, \&
  {Croton}}]{Seiler2019}
{Seiler} J., {Hutter} A., {Sinha} M., {Croton} D., 2019, \mnras, 487, 5739

\bibitem[{{Shapiro} {et~al}\mbox{.}(2004){Shapiro}, {Iliev}, \&
  {Raga}}]{shapiro2004}
{Shapiro} P.~R., {Iliev} I.~T., {Raga} A.~C., 2004, \mnras, 348, 753

\bibitem[{{Sharma} {et~al}\mbox{.}(2016){Sharma}, {Theuns}, {Frenk}, {Bower},
  {Crain}, {Schaller}, \& {Schaye}}]{sharma2016}
{Sharma} M., {Theuns} T., {Frenk} C., {Bower} R., {Crain} R., {Schaller} M.,
  {Schaye} J., 2016, \mnras, 458, L94

\bibitem[{{Sobacchi} \& {Mesinger}(2013)}]{sobacchi2013a}
{Sobacchi} E., {Mesinger} A., 2013, \mnras, 432, L51

\bibitem[{{Springel}(2005)}]{springel2005}
{Springel} V., 2005, \mnras, 364, 1105

\bibitem[{{Trebitsch} {et~al}\mbox{.}(2018){Trebitsch}, {Volonteri}, {Dubois},
  \& {Madau}}]{Trebitsch2018}
{Trebitsch} M., {Volonteri} M., {Dubois} Y., {Madau} P., 2018, \mnras, 478,
  5607

\bibitem[{{Whalen} {et~al}\mbox{.}(2004){Whalen}, {Abel}, \&
  {Norman}}]{whalen2004}
{Whalen} D., {Abel} T., {Norman} M.~L., 2004, \apj, 610, 14

\bibitem[{{Williams} {et~al}\mbox{.}(2018){Williams}, {Curtis-Lake},
  {Hainline}, {Chevallard}, {Robertson}, {Charlot}, {Endsley}, {Stark},
  {Willmer}, {Alberts}, {Amorin}, {Arribas}, {Baum}, {Bunker}, {Carniani},
  {Crand all}, {Egami}, {Eisenstein}, {Ferruit}, {Husemann}, {Maseda},
  {Maiolino}, {Rawle}, {Rieke}, {Smit}, {Tacchella}, \&
  {Willott}}]{williams2018}
{Williams} C.~C. {et~al.}, 2018, \apjs, 236, 33

\bibitem[{{Wise} \& {Cen}(2009)}]{wise2009}
{Wise} J.~H., {Cen} R., 2009, \apj, 693, 984

\bibitem[{{Wise} {et~al}\mbox{.}(2014){Wise}, {Demchenko}, {Halicek}, {Norman},
  {Turk}, {Abel}, \& {Smith}}]{wise2014}
{Wise} J.~H., {Demchenko} V.~G., {Halicek} M.~T., {Norman} M.~L., {Turk} M.~J.,
  {Abel} T., {Smith} B.~D., 2014, \mnras, 442, 2560

\bibitem[{{Yajima} {et~al}\mbox{.}(2011){Yajima}, {Choi}, \&
  {Nagamine}}]{yajima2011}
{Yajima} H., {Choi} J.-H., {Nagamine} K., 2011, \mnras, 412, 411

\bibitem[{{Yung} {et~al}\mbox{.}(2019){Yung}, {Somerville}, {Finkelstein},
  {Popping}, \& {Dav{\'e}}}]{yung2019}
{Yung} L.~Y.~A., {Somerville} R.~S., {Finkelstein} S.~L., {Popping} G.,
  {Dav{\'e}} R., 2019, \mnras, 483, 2983

\bibitem[{{Yung} {et~al}\mbox{.}(2020){Yung}, {Somerville}, {Finkelstein},
  {Popping}, {Dav{\'e}}, {Venkatesan}, {Behroozi}, \& {Ferguson}}]{yung2020}
{Yung} L.~Y.~A., {Somerville} R.~S., {Finkelstein} S.~L., {Popping} G.,
  {Dav{\'e}} R., {Venkatesan} A., {Behroozi} P., {Ferguson} H.~C., 2020,
  \mnras, 496, 4574

\end{thebibliography}

\appendix

\section{Environment-dependent galaxy number density}
\label{app_ngal}

\begin{figure}
    \centering
    \includegraphics[width=0.5\textwidth]{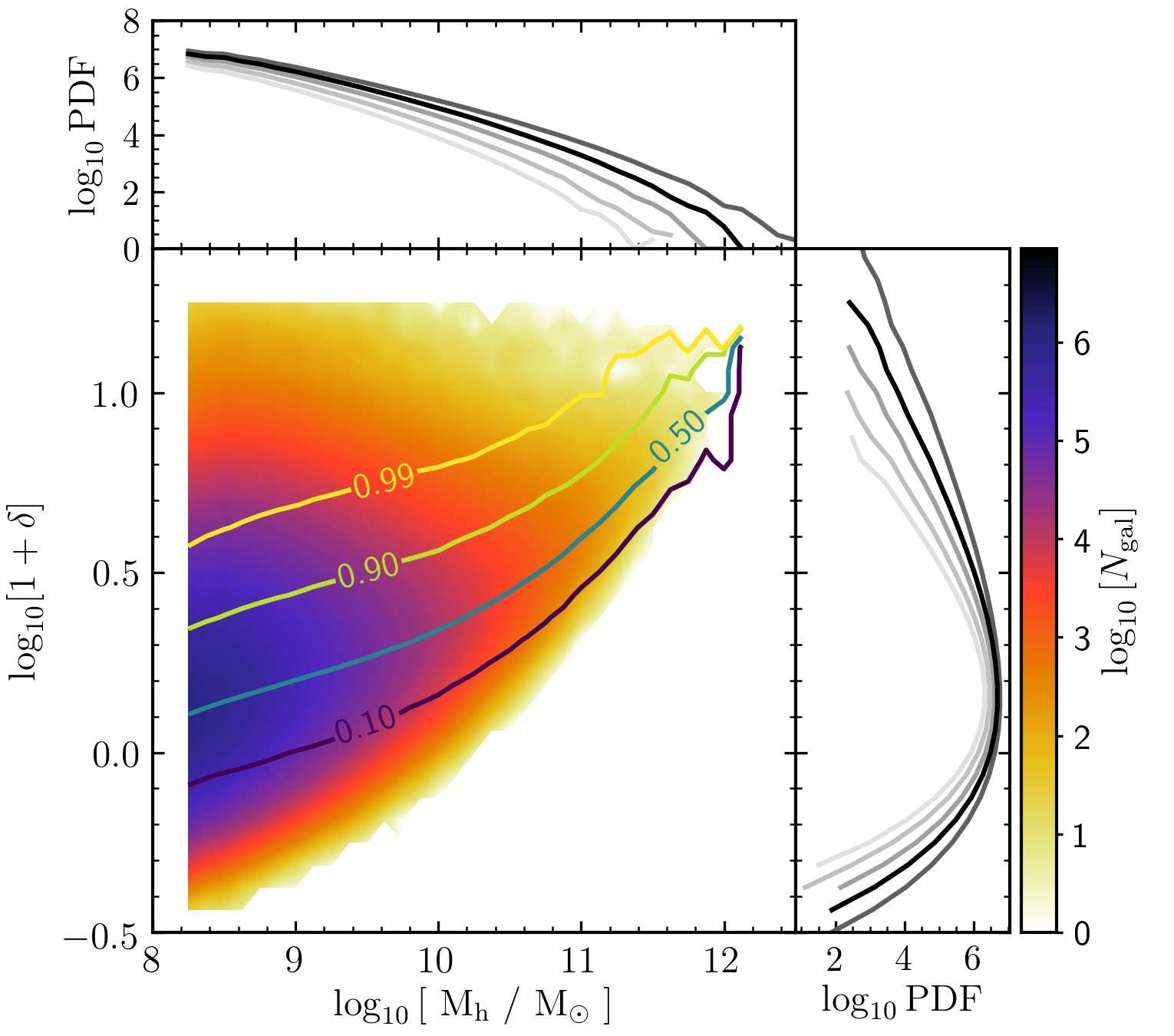}
    \caption{The number of galaxies as a function of halo mass and the over-density smoothed over $2$~cMpc at $z=7$ is shown for all models in the bottom left panel. The top left (bottom right) depicts the 1D distribution of the number of galaxies as a function of halo mass (over-density) at $z=7$. The over-plotted coloured horizontal contour lines indicate which fraction of galaxies with a halo mass $M_h$ are located in regions less dense than indicated by the contour line.}
    \label{fig_Ngal_dens_Mvir_SOBACCHI}
\end{figure}

We briefly discuss the number density distribution of galaxies in the {\sc vsmdpl} simulation in terms of their mass and environment (over-density) by means of Fig. \ref{fig_Ngal_dens_Mvir_SOBACCHI}. In this Figure, the coloured contours in bottom left panel show the number of galaxies in our simulation box as a function of the halo mass ($M_h$) and the underlying over-density ($1+\delta$) smoothed over $2$~cMpc scales. As expected from hierarchical galaxy formation, low-mass galaxies in weakly over-dense regions ($\log_{10}(1+\delta)\simeq0.1-0.2$) are the most abundant population. Galaxies with higher halo masses become increasingly less abundant and are located in increasingly over-dense regions.
The horizontal coloured lines in Fig. \ref{fig_Ngal_dens_Mvir_SOBACCHI} indicate the underlying density below which the indicated fraction of galaxies of a halo mass $M_h$ resides. We find the horizontal contour line, that marks where $50$\% of the galaxies reside in regions with over-densities lower than indicated by the line, to be constant across all redshifts ($z=5-10$), while those marking lower ($10$\%) and higher ($90$\%) thresholds move towards lower and higher densities respectively as redshift decreases. This evolution is in agreement with hierarchical structure formation where gravity causes the formation of increasingly massive objects and denser confined regions with time, while initially slightly under-dense regions loose more and more matter to over-dense regions, leading to a decrease in their matter density.

\section{The impact of low-mass halos on reionization}
\label{app_reion_contribution_minihalos}

\begin{figure}
    \centering
    \includegraphics[width=0.5\textwidth]{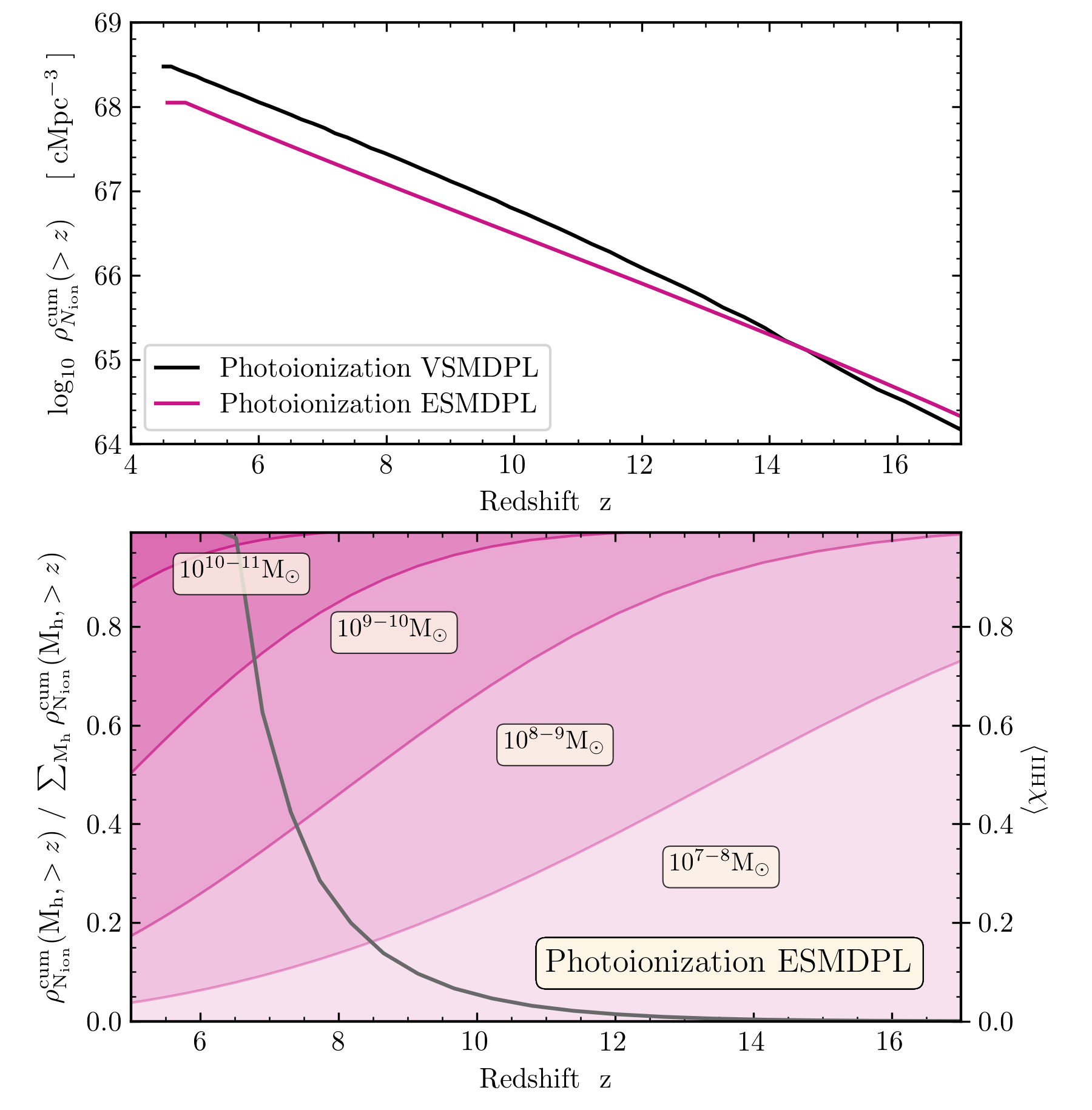}
    \caption{Top: Evolution of the density of the total number of ionizing photons at redshift $z$ for the {\it Photoionization} model and {\sc esmdpl} (violet) and {\sc vsmdpl} (black) simulation. Bottom: Fraction of the ionizing photon contribution to reionization from galaxies with halo masses within the indicated range at redshift $z$ for the {\sc esmdpl} {\it Photoionization} model. From light to darker shades the halo mass increases. The gray solid line indicates the evolution of the volume-averaged hydrogen ionization fraction.}
    \label{fig_cumNion_binnedMvir_ESMDPL}
\end{figure}

In order to estimate the number of ionizing photons originating from halos below the resolution limit of the {\sc vsmdpl} simulation ($M_h\lesssim10^{8.2}\msun$) discussed in this paper, we use the {Extremely Small MultiDark Planck} ({\sc esmdpl}) simulation that has a $20\times$ better mass resolution. The {\sc esmdpl} simulation has a box size of $64h^{-1}$~cMpc and contains $4096^3$ particles, resulting in a DM particle mass of $3.3\times10^5h^{-1}\msun$. It assumes the same cosmological parameters as the {\sc vsmdpl} simulation, $[\Omega_\Lambda, \Omega_m, \Omega_b, h, n_s, \sigma_8]=[0.69, 0.31, 0.048, 0.68, 0.96, 0.83]$. As for the {\sc vsmdpl} simulation, we identify all halos and subhalos with at least $20$ particles using the phase-space halo finder {\sc rockstar} \citep{behroozi2013_rs} in each snapshot, and build the corresponding merger trees using the {\sc consistent trees} method \citep{behroozi2013_trees}. Finally, we resort the resulting vertical merger trees using {\sc cutnresort} within the {\sc astraeus} pipeline to a local horizontal layout. We run our {\sc astraeus} on the local horizontal trees and density fields ($256^3$ grids) of the {\sc esmdpl} simulation assuming the {\it Photoionization} radiative feedback model.

The corresponding evolution of the cumulative number density of ionizing photons until redshift $z$ (top) and the contribution of different halo masses to the ionizing budget from $z\simeq25$ down to redshift $z$ (bottom) are shown in Fig. \ref{fig_cumNion_binnedMvir_ESMDPL}. 
Firstly, from the bottom panel, we see that over the course of reionization very low-mass halos ($M_h=10^7-10^8\msun$) contribute less than $7\%$ of the ionizing budget until $z=6$. They dominate the ionizing budget only at higher redshifts, $z\gtrsim14$, when reionization has barely started. 
Secondly, we note that the $10^{8.2-9}\msun$ halos in the {\sc vsmdpl} simulation contribute a similar amount of ionizing photons to reionization as the $10^{8-9}\msun$ halos in the {\sc esmdpl} simulation. This agreement can be explained by the initial starburst in the $M_h\simeq10^{8.2}\msun$ halos in the {\sc vsmdpl} simulation that produce more ionizing photons than those in the {\sc esmdpl} simulation \citep[see star formation rates in Fig. B1 and B2 in][]{hutter2021}.
Finally, the similar total ionizing emissivities in the {\sc vsmpdl} and {\sc esmdpl} simulations (c.f. black and violet lines in top panel of Fig. \ref{fig_cumNion_binnedMvir_ESMDPL}) are also reflected in their nearly identical values of the ionizing escape fraction $f_\mathrm{esc}$ ($24\%$ and $21.5\%$ for the {\sc esmdpl} and {\sc vsmdpl} simulations, respectively) that are required to reproduce the optical depth measured by Planck.

\section{Evolution of the instantaneous ionizing emissivity}
\label{app_ionizing_emissivity}

\begin{figure}
    \centering
    \includegraphics[width=0.47\textwidth]{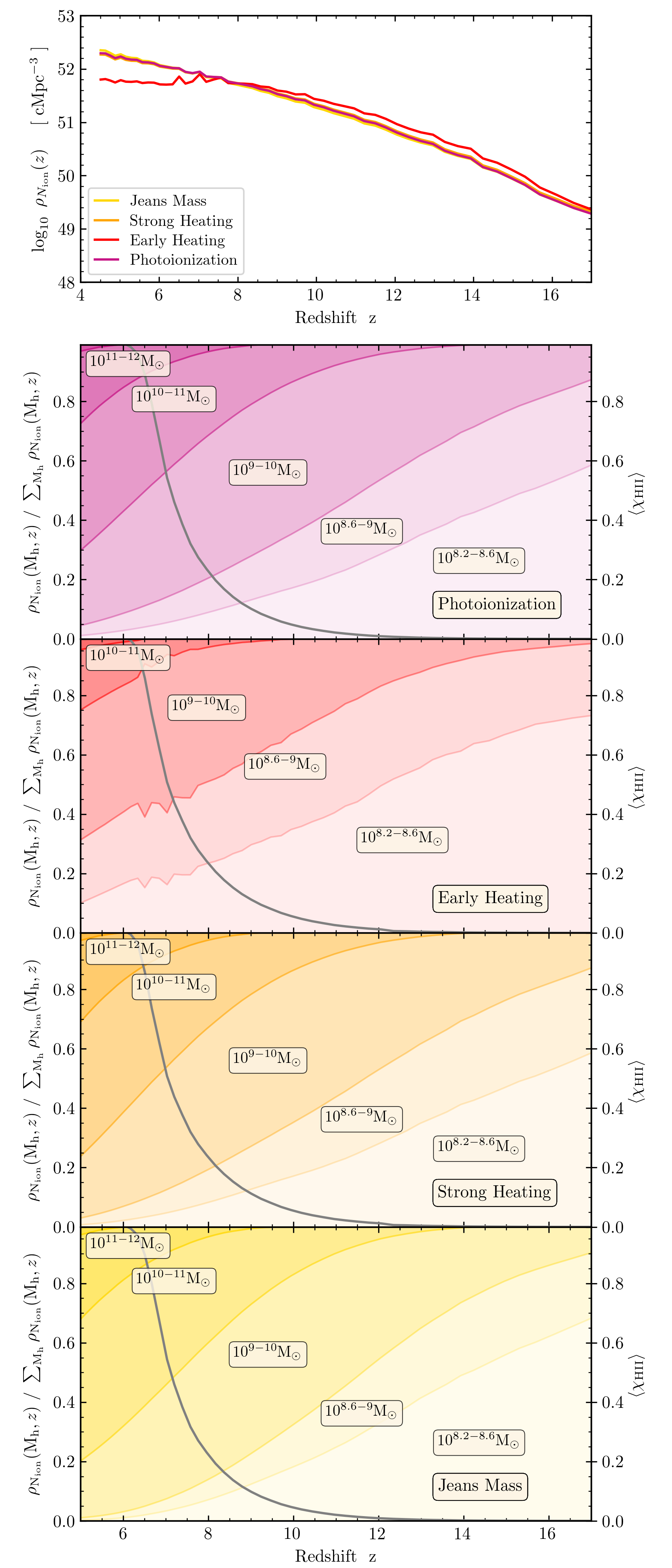}
    \caption{Top: Evolution of the density of the total number of ionizing photons at redshift $z$ for the {\it Photoionization} (violet), {\it Early Heating} (red), {\it Strong Heating} (orange) and {\it Jeans Mass} model (yellow). Bottom: Fraction of the ionizing photon contribution to reionization from galaxies with halo masses within the indicated range at redshift $z$ for the {\it Photoionization} (violet), {\it Early Heating} (red), {\it Strong Heating} (orange) and {\it Jeans Mass} model (yellow). From light to darker shades the halo mass increases. The gray solid line indicates the evolution of the volume-averaged hydrogen ionization fraction.}
    \label{fig_Nion_binnedMvir_contribution}
\end{figure}

In order to enable a direct comparison with other works, we show also the relative instantaneous contribution of ionizing photons of different halo masses throughout reionization in Fig. \ref{fig_Nion_binnedMvir_contribution}. Similar to the results in Section \ref{sec_ionizing_budget}, we see that in models with constant $f_\mathrm{esc}$ values low-mass galaxies ($M_h\lesssim10^9\msun$) contribute the majority of ionizing photons before reionization starts at $z\lesssim10$. Their contribution drops then to about $<10$\% at the midpoint of reionization ($z=7$). Scaling the ionizing escape fraction with the fraction of gas that is ejected from the galaxy, i.e. leading to a decreasing $f_\mathrm{esc}$ with halo mass $M_h$, results in a far more significant contribution of low-mass halos providing about $\sim45$\% of ionizing photons at $z=7$.

\subsection{The impact of non-continuous time steps on $f_\mathrm{esc}^\mathrm{ej}$}
\label{app_subsec_oscillations}

From Fig. \ref{fig_Nion_binnedMvir_contribution} we note that the instantaneous total ionizing emissivity oscillates between $z\simeq7.5$ and $z\simeq6$ in the {\it Early Heating} model. These oscillations are directly linked to the oscillations of the ionizing escape fraction ($f_\mathrm{esc}^\mathrm{ej}$) and thereby the ejected gas fraction ($f_\star^\mathrm{eff}/f_\star^\mathrm{ej}$, see equation \ref{eq_fej_delayedSN} for $f_\star^\mathrm{ej}$), which again are caused by the discreteness of our delayed SN feedback scheme in combination with the time span of the time steps at $z=7.5-6$. In order to understand this interplay, we can consider equation \ref{eq_fej_delayedSN}. From this equation, we can see that $f_\star^\mathrm{ej}$ increases firstly as the time step approaches the lifetime of a $8\msun$ star that explodes as a SN ($28.6$~Myr), and secondly as the ratio between star formation in previous time steps and current (initial) gas mass decreases. In general, the underlying merger trees have time steps that scale with the logarithm of the scale factor, however, in order to cover some redshifts where the majority of observations are performed, they have been minorly adjusted. Hence, between $z=8-6$ not all time steps are necessarily larger than the previous ones, a few (in total $3$) are $0.8-1.7$~Myr shorter than their previous ones. In a time step that is shorter than the previous one, $\nu_j$ in equation \ref{eq_fej_delayedSN} does not decrease compared to the previous time step but increases, resulting in lower $f_\star^\mathrm{ej}$ values and less stars being formed (minimum). In the subsequent time step both $\nu_j$ and $M_{\star,j}^\mathrm{new}$ decrease again, leading to the expected increase in $f_\star^\mathrm{ej}$ (maximum) and the formation of more stars. The reduced star formation again flattens the increase of $f_\star^\mathrm{ej}$ in the then subsequent time step. At $z\simeq6.5$ the length of the time steps surpasses the lifetime of a $8\msun$ star and hence our delayed SN feedback scheme converges to the instantaneous SN feedback scheme, where $f_\star^\mathrm{ej}$ is solely sensitive to the gravitational potential of the underlying halo.

\label{lastpage} 
\end{document}